\documentclass[10pt, conference, letterpaper]{IEEEtran}
\IEEEoverridecommandlockouts

\usepackage[
  backend=biber,
  style=ieee,
  minbibnames=100, maxbibnames=100,
]{biblatex}
\AtBeginDocument{} 

\usepackage{multirow}
\usepackage[algo2e,linesnumbered,ruled,lined]{algorithm2e}
\SetKwInOut{Input}{Input}
\SetKwInOut{Output}{Output}
\SetKwComment{Comment}{/* }{ */}

\SetAlFnt{\small\sffamily}

\SetKw{Continue}{continue}
\usepackage{array}
\usepackage{multirow}
\usepackage[table,xcdraw]{xcolor}
\usepackage{stfloats}

\newcolumntype{P}[1]{>{\centering\arraybackslash}p{#1}}

\makeatletter
\def\Cline#1#2{\@Cline#1#2\@nil}
\def\@Cline#1-#2#3\@nil{%
  \omit
  \@multicnt#1%
  \advance\@multispan\m@ne
  \ifnum\@multicnt=\@ne\@firstofone{&\omit}\fi
  \@multicnt#2%
  \advance\@multicnt-#1%
  \advance\@multispan\@ne
  \leaders\hrule\@height#3\hfill
  \cr}
\makeatother

\usepackage{amsmath,amssymb,amsfonts}
\usepackage{graphicx}
\usepackage{subcaption} 
\usepackage{textcomp}
\usepackage{xcolor} 
\usepackage[hidelinks]{hyperref}
\usepackage{booktabs,threeparttable}

\usepackage{tabularx}
\usepackage{tikz}
\usepackage{pgfplots}
\pgfplotsset{compat=1.18}
\usetikzlibrary{pgfplots.statistics}
\usetikzlibrary{patterns}
\usepackage{amsfonts}
\usepackage[capitalise]{cleveref}

\usepackage{stfloats}

\usepackage{framed} 

\usepackage[acronym]{glossaries}
\newacronym{dag}{DAG}{Directed Acyclic Graph}
\newacronym{qor}{QoR}{Quality of Result}
\newacronym{qos}{QoS}{Quality of Service}
\newacronym{qoe}{QoE}{Quality of Experience}
\newacronym{yolo}{YOLO}{You Only Look Once}
\newacronym{rcnn}{R-CNN}{Region-based Convolutional Neural Network}
\newacronym{hog}{HOG}{Histograms of Oriented Gradients}
\newacronym{cnn}{CNN}{Convolutional Neural Network}
\newacronym{iot}{IoT}{Internet of Things}
\newacronym{iiot}{IIoT}{Industrial Internet of Things}
\newacronym[plural=FMs,firstplural=Feature Models]{fm}{FM}{Feature Model}
\newacronym[plural=SPLs,firstplural=Software Product Lines]{spl}{SPL}{Software Product Line}
\newacronym{JMH}{JMH}{Java Microbenchmark Harness}
\newacronym{cma}{CMA}{Cumulative Moving Average}
\newacronym[plural=CTCs,firstplural=Cross-Tree Constraints]{ctc}{CTC}{Cross-Tree Constraint}
\newacronym{dspl}{DSPL}{Dynamic Software Product Line}
\newacronym[plural=EFMs,firstplural=Extended Feature Models]{efm}{EFM}{Extended Feature Model}
\newacronym[plural=PFMs,firstplural=Parial Feature Models]{pfm}{PFM}{Partial Feature Model}
\newacronym{cnf}{CNF}{Conjunctive Normal Form}
\newacronym[plural=NFVs,firstplural=Network Functions Virtualizations]{nfv}{NFV}{Network Functions Virtualization}
\newacronym[plural=RQs,firstplural=Research Questions]{rq}{RQ}{Research Question}
\newacronym{v2x}{V2X}{Vehicle-to-Everything}
\newacronym{sat}{SAT}{Boolean Satisfiability Problem}
\newacronym[plural=NPUs,firstplural=Neural Processing Units]{npu}{NPU}{Neural Processing Unit}
\newacronym{RAPL}{RAPL}{Running Average Power Limit}
\newacronym{Kepler}{Kepler}{Kubernetes-based Efficient Power Level Exporter}
\newacronym{ocr}{OCR}{Optical Character Recognition}
\newacronym{ALPR}{ALPR}{Automatic License Plate Recognition}
\newacronym{SVR}{SVR}{Support Vector Regression}
\newacronym{MAE}{MAE}{Mean Absolute Error}
\newacronym{ETL}{ETL}{Extract–Transform–Load}
\newacronym{MAPE}{MAPE}{Mean Absolute Percentage Error}
\newacronym{OHE}{OHE}{one-hot encoder}
\newacronym[plural=SLOs,firstplural=service-level objectives]{SLO}{SLO}{service-level objective}
\newacronym{CRI}{CRI}{Container Runtime Interface}
\newacronym{IQR}{IQR}{interquartile ranges}
\newacronym{mAP}{mAP}{Mean Average Precision}
\newacronym{RMSE}{RMSE}{Root Mean Squared Error}
\newacronym[plural=MLPs,firstplural=Multilayer Perceptrons]{mlp}{MLP}{Multilayer Perceptron}
\newacronym{FCFS}{FCFS}{First-Come-First-Serve}
\newacronym{MEC}{MEC}{Mobile Edge Computing}
\newacronym{WTP}{WTP}{willingness to pay}

\definecolor{rowgray}{gray}{0.85} 

\usepackage{flushend}
\usepackage{scalerel}
\usetikzlibrary{svg.path}

\definecolor{orcidlogocol}{HTML}{A6CE39}
\tikzset{
  orcidlogo/.pic={
    \fill[orcidlogocol] svg{M256,128c0,70.7-57.3,128-128,128C57.3,256,0,198.7,0,128C0,57.3,57.3,0,128,0C198.7,0,256,57.3,256,128z};
    \fill[white] svg{M86.3,186.2H70.9V79.1h15.4v48.4V186.2z}
                 svg{M108.9,79.1h41.6c39.6,0,57,28.3,57,53.6c0,27.5-21.5,53.6-56.8,53.6h-41.8V79.1z M124.3,172.4h24.5c34.9,0,42.9-26.5,42.9-39.7c0-21.5-13.7-39.7-43.7-39.7h-23.7V172.4z}
                 svg{M88.7,56.8c0,5.5-4.5,10.1-10.1,10.1c-5.6,0-10.1-4.6-10.1-10.1c0-5.6,4.5-10.1,10.1-10.1C84.2,46.7,88.7,51.3,88.7,56.8z};
  }
}

\newcommand\orcidicon[1]{\href{https://orcid.org/#1}{\mbox{\scalerel*{
\begin{tikzpicture}[yscale=-1,transform shape]
\pic{orcidlogo};
\end{tikzpicture}
}{|}}}}
\usepackage{balance}

\makeatletter
\newcommand{\copyrightnotice}[1]{\gdef\@copyrightnotice{#1}}
\newcommand{\toappear}[1]{\gdef\@toappear{#1}}
\let\@copyrightnotice\@empty
\let\@toappear\@empty
\newcommand{\ps@myfooter}{%
  \let\@mkboth\@gobbletwo
  \def\@oddhead{}%
  \def\@evenhead{}%
  \def\@oddfoot{\rlap{\@toappear}\hfil\thepage\hfil\llap{\@copyrightnotice}}%
  \let\@evenfoot\@oddfoot
}
\makeatother

\toappear{\it To appear in Proc.\ WACV2025}
\copyrightnotice{\copyright\ 2025 IEEE}
\usepackage{ifthen}
\usepackage{atbegshi}
\newcounter{pagecounter}
\AtBeginShipout{%
  \stepcounter{pagecounter}%
  \ifthenelse{\value{pagecounter}=1}{%
    \AtBeginShipoutUpperLeft{%
      \put(\dimexpr\paperwidth-20.3cm\relax,-0.8cm){%
        \parbox[t]{19cm}{%
          \copyright\ 2026 IEEE. Personal use of this material is permitted. 
          Permission from IEEE must be obtained for all other uses, in any current or future media, including reprinting/republishing this material for advertising or promotional purposes, creating new collective works, for resale or redistribution to servers or lists, or reuse of any copyrighted component of this work in other works.%
        }%
      }%
    }%
  }{}%
}

\begin{document}

\title{PRICE: Pricing-based Resource Incentives for Quality-of-Result-aware Computing at the Edge}

\author{
    \IEEEauthorblockN{
        Uwe Gropengießer%
        \IEEEauthorrefmark{1}\orcidicon{0000-0002-1334-8538},
        Sebastian Frenz%
        \IEEEauthorrefmark{2}\orcidicon{0009-0007-7046-1733},
        Max \discretionary{Mühl-}{häuser}{Mühlhäuser}%
        \IEEEauthorrefmark{3}\orcidicon{0000-0003-4713-5327},
    }
    \IEEEauthorblockA{
        \IEEEauthorrefmark{1}
        \IEEEauthorrefmark{2}
        \IEEEauthorrefmark{3}
        Technical University of Darmstadt; Darmstadt, Germany\\
        Email:%
        \IEEEauthorrefmark{1}uwe.gropengiesser@tu-darmstadt.de,
        \IEEEauthorrefmark{2}sebastian.daniel.frenz@gmail.com,
        \IEEEauthorrefmark{3}max@informatik.tu-darmstadt.de
    }
}

\maketitle
\thispagestyle{plain}
\pagestyle{plain}

\begin{abstract}
Edge nodes are capacity-constrained by design, yet many edge workloads can trade result quality for resource efficiency at runtime. Existing edge pricing mechanisms largely treat requests as fixed-configuration submissions and rarely exploit per-request quality flexibility under overload. We present \textsc{PRICE}, an incentive mechanism that couples a utilization-dependent price signal to per-request quality selection. As utilization increases, rising acceptance prices make resource-intensive variants less likely to be selected, shifting accepted requests toward lighter execution and allowing the node to serve significantly more requests while operating near capacity. Evaluation on real hardware under sustained overload shows that \textsc{PRICE} outperforms both fixed-allocation and dynamic-pricing baselines in accepted throughput and CPU utilization. The results are robust across pricing function families, task-duration distributions, and client populations. Result-quality flexibility is a powerful but underused control dimension for overload management at the edge, and pricing is an effective mechanism to exploit it.
\end{abstract}


\glsresetall
\section{Introduction}\label{sec:Introduction}

Edge computing places computation close to data sources and end users, enabling latency-sensitive applications that cloud deployments cannot serve within tight time budgets~\cite{shi_edge_2016, satyanarayanan_emergence_2017, charyyev_latency_2020}. Unlike cloud data centers, edge nodes operate under tight computational and memory constraints~\cite{satyanarayanan_emergence_2017, gedeon_what_2019, pradhan_towards_2024, gropengieser_marq_2025}, and when multiple tenants concurrently offload tasks to the same node, demand can exceed available capacity. The applications driving edge adoption, including autonomous driving~\cite{grigorescu_survey_2020}, industrial monitoring~\cite{ngo_adaptive_2021}, and augmented reality~\cite{gedeon_what_2019}, share infrastructure in multi-tenant deployments and submit requests on demand. Independent tenants generate bursty and heterogeneous workloads without coordinating with each other or with the provider, making provider-side resource allocation under dynamic load a central systems challenge.

Early approaches relied on baseline scheduling assumptions such as FCFS in M/M/c queueing models~\cite{jia_optimal_2017} or static offloading schemes with fixed decisions~\cite{zhang_offloading_2015}. These are straightforward to implement but offer no control dimension beyond queueing and rejection once demand approaches capacity. Under fluctuating load, queue buildup leads to unstable response times and wasted resources, while static allocation is unable to adapt to the bursty and heterogeneous workloads typical of multi-tenant edge deployments. To cope with these limitations, recent work has introduced economic and game-theoretic mechanisms, including multichoice games with Shapley-value-based cost sharing~\cite{li_multichoice_2018}, cyclic games studying mixed cooperation and competition~\cite{ma_cyclic_2019}, non-cooperative multidimensional allocation~\cite{ye_non-cooperative_2013}, cooperative formulations for fairness in virtualized infrastructures~\cite{kim_cooperative_2020}, and market-based pricing mechanisms~\cite{park_real-time_2024}. Pricing is particularly attractive because it enables decentralized self-selection among competing tenants without requiring explicit coordination~\cite{kelly_rate_1998, park_real-time_2024}. However, these mechanisms treat each request as a fixed-configuration submission, leaving acceptance or rejection as the only control levers regardless of their pricing sophistication.

What these approaches overlook is that many edge workloads can produce results at multiple quality levels with substantially different resource demands. Applications such as object detection~\cite{Li2017, gropengieser_marq_2025}, video analytics~\cite{wen_approxiot_2018, pandey_mobidic_2016}, and sensor fusion~\cite{ngo_adaptive_2021, younis_qlran_2021} expose this flexibility at runtime, as do data analytics pipelines more broadly~\cite{goiri_approxhadoop_2015, quoc_streamapprox_2017, hu_approximation_2019}. These workloads expose multiple execution variants that differ in computational demand and achievable result quality. Critically, many of these applications are latency-sensitive and cannot tolerate indefinite queueing or multi-round bidding. Clients must be able to fall back to local execution when offloading fails. PRICE supports this by rejecting non-feasible requests immediately rather than queueing them, preserving predictable response behavior under overload. Ignoring result-quality flexibility therefore foregoes a powerful demand-shaping mechanism that would allow the system to serve more requests immediately rather than deferring or rejecting them outright.

We present \emph{PRICE} (Pricing-based Resource Incentives for Quality-of-Result-aware Computing at the Edge), a pricing-based incentive mechanism that couples a utilization-dependent price signal to per-request quality selection. As utilization increases, rising acceptance prices make resource-intensive variants less likely to be selected, while falling utilization makes higher-quality variants reachable again. This self-regulating behavior drives the system toward sustained high utilization near capacity, improves effective throughput, and exposes result quality as an adaptive control dimension under contention, without requiring tenant coordination, workload models, or prior knowledge of tenant preferences.

Following a motivating example (\cref{sec:Background}), \cref{sec:ProblemFormulation} formalizes the model and incentive perspective. \cref{sec:SystemDesign} and \cref{sec:Approach} present the system architecture and pricing mechanism. \cref{sec:Evaluation} evaluates PRICE, followed by discussion, related work, and conclusion.
\begin{figure*}[ht!]
    \centering
    \includegraphics[width=0.8\linewidth]{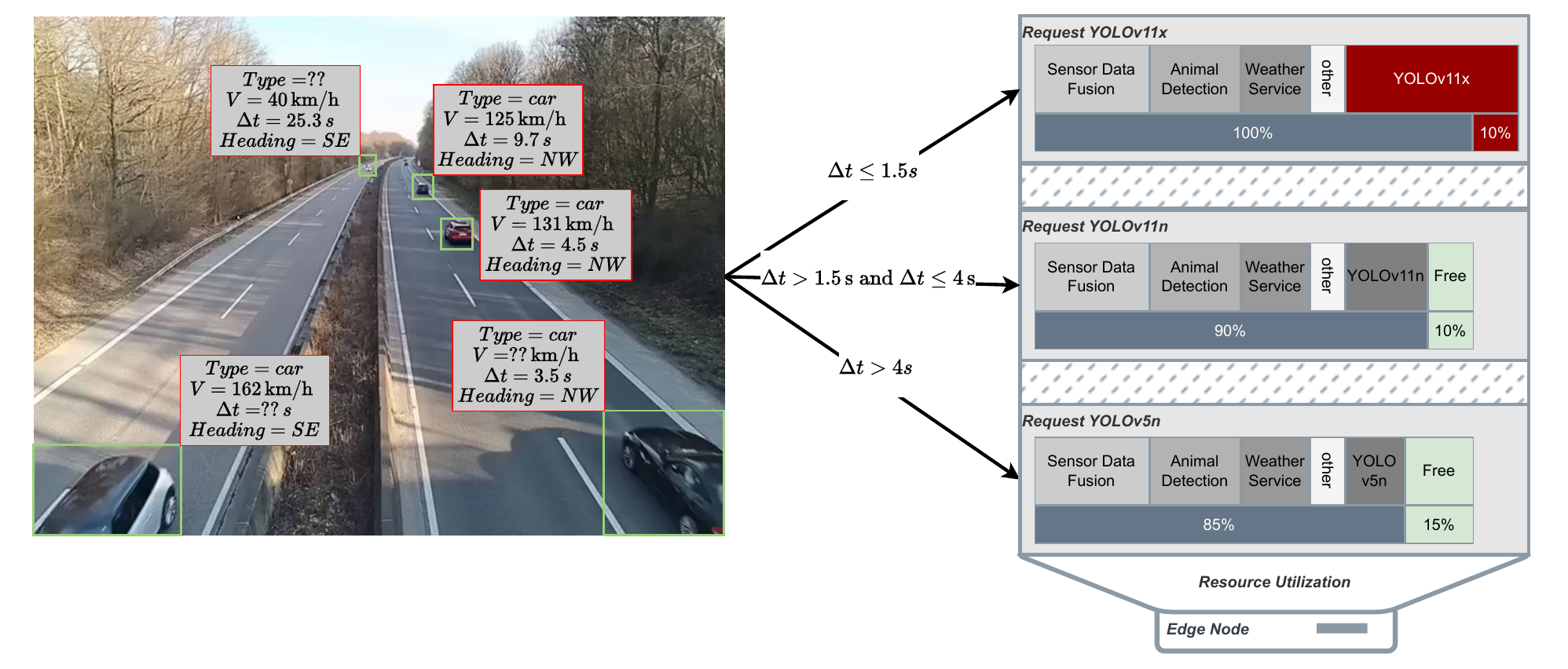}
    \caption{Motivating example for QoR-aware variant selection at the edge. A traffic congestion detection service switches among three execution variants based on the inter-vehicle time gap $\Delta t$. Sparse traffic ($\Delta t > 4\,\text{s}$) allows a lightweight model with low resource demand, while dense traffic ($\Delta t \leq 1.5\, \text{s}$) requires a more accurate, compute-intensive variant. Under high load at the edge node, resource-intensive variants may exceed capacity, illustrating the trade-off between result quality, resource consumption, and available capacity that PRICE addresses through utilization-dependent pricing.}   
    \label{fig:MotivatingExample}
\end{figure*}
\section{Background and Motivating Example}\label{sec:Background}

Our work builds on the observation that many edge workloads can trade result quality for resource efficiency at runtime. We capture this trade-off through \gls{qor}, a normalized quality metric that abstracts domain-specific measures into a common scale. A \gls{qor} of 100\,\% denotes the highest attainable result quality, while 0\,\% corresponds to complete failure~\cite{gropengieser_marq_2025}. Domain-specific quality metrics can be mapped to this scale. For example, in computer vision, \gls{qor} is often expressed through metrics such as \gls{mAP} for object detection~\cite{padilla_survey_2020} or word-level accuracy for text recognition~\cite{baek_what_2019}. In contrast, relative error is a common measure in data analytics~\cite{goiri_approxhadoop_2015, wen_approxiot_2018, hu_approximation_2019}. In some domains, quality can also be estimated at runtime without ground truth, for example, through approximate analytics techniques in data processing pipelines~\cite{goiri_approxhadoop_2015, quoc_streamapprox_2017, wen_approxiot_2018}. This abstraction enables a unified allocation model in which heterogeneous tenants express quality requirements in a comparable form.

To place \gls{qor} in a broader systems context, we build on approximate computing, which treats result quality as a controllable variable for resource efficiency. Approximate computing provides the fundamental paradigm for deliberately varying \gls{qor} to achieve substantial savings in resource usage, execution time, and energy consumption. It describes the controlled reduction of computational accuracy when the resulting quality loss is compensated by efficiency gains~\cite{mittal_survey_2016, leon_approximate_2025, liu_approximate_2020}. In this paper, \gls{qor} serves as a normalized percentage-based indicator that enables domain-independent comparability across heterogeneous workloads~\cite{gropengieser_marq_2025, pandey_mobidic_2016}. The key insight is that many edge applications already expose this flexibility through discrete execution variants~\cite{Gedeon2021}, and that a provider can exploit it as a control dimension for resource management.

We use the following video analytics service as a running example, where the application can switch among discrete execution variants that differ in resource demand and achievable \gls{qor} (\cref{fig:MotivatingExample}). This example serves as a conceptual illustration throughout the paper and is not tied to a specific model or dataset. The service detects traffic congestion based on continuous video analysis and estimates the temporal gap between vehicles, denoted by $\Delta t$. As $\Delta t$ decreases, traffic density increases, making timely detection more important, which motivates higher sampling rates and more accurate detection methods. When $\Delta t$ is larger, the service can reduce sampling and switch to a less complex variant to save resources. Under dense traffic, visual occlusion increases, and robust detection often requires more compute-intensive models. When a resource-intensive variant is requested and the node is already under load, accepting it may exhaust available capacity, leaving no headroom for other tenants or subsequent requests. For congestion detection, however, fine-grained vehicle classification is not required. The key objective is to identify congestion early enough to trigger warnings or rerouting. It is therefore often sufficient to run a cheaper variant that yields a lower \gls{qor} but remains acceptable under resource scarcity. Such runtime switching among variants has been demonstrated in prior systems~\cite{gropengieser_marq_2025, pandey_mobidic_2016}. We assume that tasks request resources on demand rather than continuously occupying them.

The same pattern appears in a wide range of edge applications, and the flexibility it offers is substantial. Object recognition and media processing applications expose control knobs that allow trading recognition accuracy for reduced processing time and energy consumption~\cite{Li2017, pandey_mobidic_2016, gropengieser_marq_2025}. Industrial monitoring pipelines balance detection accuracy against processing latency under resource constraints~\cite{wen_approxiot_2018, ngo_adaptive_2021}, while perception pipelines in connected vehicles trade inference accuracy for computational efficiency to meet real-time constraints~\cite{grigorescu_survey_2020}. Across these settings, applications expose multiple execution variants that differ in achievable \gls{qor}, runtime, and resource demand, and the gap between the most and least resource-intensive variants can be substantial.

The challenge becomes pronounced on shared edge nodes hosting multiple independent tenants. In a non-cooperative setting, each tenant naturally selects the variant that maximizes its own utility, which can induce correlated demand for high-\gls{qor} variants and push aggregate utilization beyond node capacity.

The resulting overload leads to longer waiting times, unstable response behavior, and inefficient use of scarce resources, particularly under stochastic arrivals.
Systems such as MARQ~\cite{gropengieser_marq_2025}, MobiDiC~\cite{pandey_mobidic_2016}, and MobiQoR~\cite{Li2017} select among execution variants within individual applications but provide no admission mechanism for competing tenants on a shared node.
A direct empirical comparison would therefore require adding such a policy, while PRICE complements this line of work by coordinating variant selection across tenants through a utilization-dependent price signal.
Tenants signal their preferences through bids, and the provider selects the feasible variant with the highest margin score among all bid-acceptable variants, steering the system toward sustained, near-capacity utilization without explicit coordination.
Throughout this paper, \emph{tenants} denote the paying parties that issue requests, \emph{clients} denote the software instances submitting them, and the \emph{provider} denotes the party that sets acceptance prices and receives the resulting revenue.
In our evaluation, each tenant is represented by a single client process.
\section{Problem Formulation}\label{sec:ProblemFormulation}

We consider an edge environment with limited computational resources, jointly used by multiple independent tenants. Tenants are indexed by $t \in \{1,\dots,T\}$. Let $I$ denote the set of resource types (e.g., CPU and RAM). Each edge node provides a capacity $c_i$ for every resource $i \in I$.

We assume that each tenant $t$ provides a finite set of discrete execution variants $\mathcal{K}_t$. A variant $k \in \mathcal{K}_t$ represents a concrete workload configuration and determines an achievable \gls{qor} $q_{t,k} \in [0,1]$ together with a required resource demand vector $R_{t,k}$ for executing the request at that quality. We denote the demand of resource $i$ by $R_{t,k}[i]$. Variants may correspond to different model sizes within the same objective (e.g., YOLO11n vs.\ YOLO11x) or to application configurations such as buffer sizes or image scaling. In the running example from \cref{fig:MotivatingExample}, variants correspond to different detection models and sampling rates used by the congestion-detection pipeline.

Since each variant requires multiple resource types, we assume that the provider exposes a single price signal $p$ that applies to the aggregate resource footprint of a request. Multi-resource demands $R_{t,k}$ are mapped to scalar resource units through an aggregation function $\rho(\cdot)$, where $\rho(R_{t,k})$ denotes the total resource footprint of variant $k$. In our realization, $\rho(\cdot)$ is not computed as an explicit preprocessing step but is realized implicitly through the per-resource marginal cost integrals described in \cref{sec:Approach}.

We further assume that tenants are willing to trade result quality for lower cost or improved execution performance, such as reduced latency or meeting application-specific deadlines. We model this trade-off through an individual utility function
\[
    U_t(q_{t,k},\, p) = V_t(q_{t,k}) - p \cdot \rho(R_{t,k}),
\]
where $V_t(q_{t,k})$ denotes the value derived from achieving the quality level $q_{t,k}$ of variant $k$, $\rho(R_{t,k})$ denotes the aggregated resource units required for that variant, and $p$ is the unit price per aggregated resource unit. We assume that $V_t(\cdot)$ is strictly increasing with diminishing returns, while $R_{t,k}$ grows nonlinearly with $q_{t,k}$.

An allocation is feasible if for every resource $i \in I$ the aggregate demand of all assigned variants does not exceed the available capacity:
\[
    \sum_{t=1}^{T} R_{t,k_t}[i] \leq c_i \quad \forall i \in I.
\]
This condition describes feasibility in terms of aggregate resource demand, without reference to the runtime utilization state $u[i]$, which is introduced in \cref{sec:Approach}.

We consider a purely non-cooperative setting in which each tenant independently aims to maximize its own utility by trading off achievable \gls{qor} against the resource prices it must pay. Tenants encode their preferences through bid values $b_{t,k}$ that reflect their willingness to pay for each variant in $\mathcal{K}_t$, as described in \cref{sec:Approach}. 

This non-cooperative decision-making can lead to an over-demand situation in which aggregate resource requests exceed capacity, resulting in overload and degraded performance for all tenants. The provider faces the challenge of adapting a utilization-dependent pricing signal to regulate demand and drive the system toward sustained high utilization near capacity. The objective is to convert available capacity into completed work as efficiently as possible while maintaining acceptable quality levels, operating close to capacity, and preserving enough headroom to absorb stochastic arrivals and bursts. Overall, the problem is to design a dynamic, quality-aware pricing mechanism that aligns individual tenant decisions with system-level efficiency under stochastic workloads and limited resources.
\section{System Design and Runtime Flow}\label{sec:SystemDesign}

PRICE is realized as a distributed edge runtime in which tenant-side clients submit tasks with multiple execution variants, while the provider coordinates placement and execution under hard resource constraints. This section introduces the system architecture, the request format, and the runtime interaction between system components. The detailed pricing and variant selection logic is described in \Cref{sec:Approach}.

\begin{figure*}[t!]
    \centering
    \includegraphics[width=0.85\linewidth]{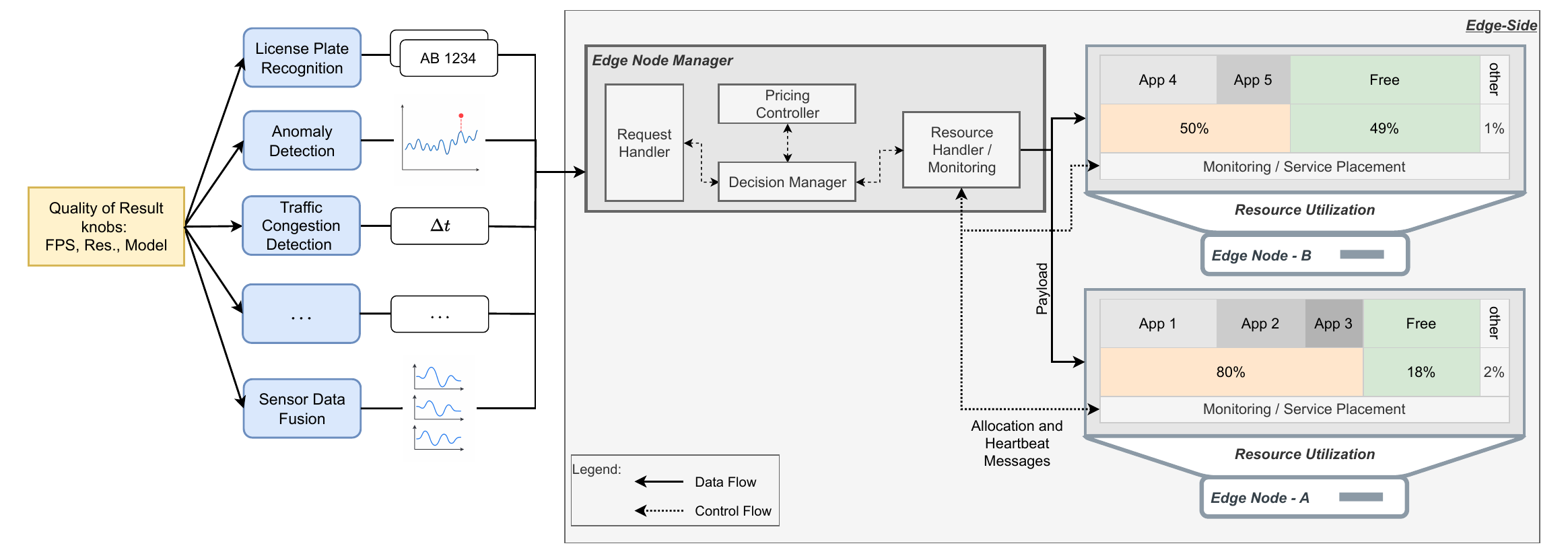}
    \caption{PRICE system architecture. Tenant-side clients submit requests containing bids over execution variants. The edge node accepts feasible and price-acceptable variants based on current utilization and capacity, while monitoring feedback drives runtime price adaptation.}
    \label{fig:SystemDesign}
    \vspace{-2mm}
\end{figure*}

\Cref{fig:SystemDesign} illustrates the PRICE architecture in a multi-tenant edge environment. Tenant applications, including the traffic congestion detection service from \cref{sec:Background}, license plate recognition, anomaly detection, and sensor fusion, expose configurable QoR knobs such as frame rate, resolution, or model choice. These configurations are represented as discrete execution variants with different resource demands and achievable QoR. Instead of submitting a single fixed configuration, each request provides multiple variants from which the system selects one at runtime. The figure distinguishes between data flow and control flow. The data flow carries payloads and results between clients and edge nodes, while the control flow comprises system coordination, including state reports, pricing updates, and scheduling decisions.

PRICE separates coordination from execution. An \textit{Edge Node Manager} maintains a global state of available edge nodes, processes incoming requests, and manages the utilization-dependent pricing signal that regulates demand. Within the Edge Node Manager, a \textit{Decision Manager} component evaluates incoming bids against the current acceptance price and selects the variant to execute, as detailed in \cref{sec:Approach}. Based on telemetry and current resource availability, the manager evaluates feasible execution variants and assigns the selected variant to a suitable node. The edge nodes execute tasks and enforce hard capacity limits by reserving and releasing resources. In our setup, applications are deployed on the nodes as containerized services (e.g., Docker containers or Kubernetes pods) ahead of time, avoiding compilation and distribution overhead at request time. PRICE does not require this deployment model and can operate with any execution environment that supports resource reservation and monitoring. Nodes continuously monitor their resource utilization and report it to the manager via periodic heartbeat messages, which the manager uses to maintain an up-to-date view of available resources across all managed nodes. The manager performs lightweight coordination only. Requests are forwarded immediately, and execution remains decentralized on the edge nodes. The design does not assume a single manager instance or a fixed number of managed nodes. Instead, multiple managers can operate concurrently, each responsible for a disjoint subset of edge nodes, allowing the system to scale with deployment size.

\begin{figure}[bt!]
    \centering
    \includegraphics[width=0.8\linewidth]{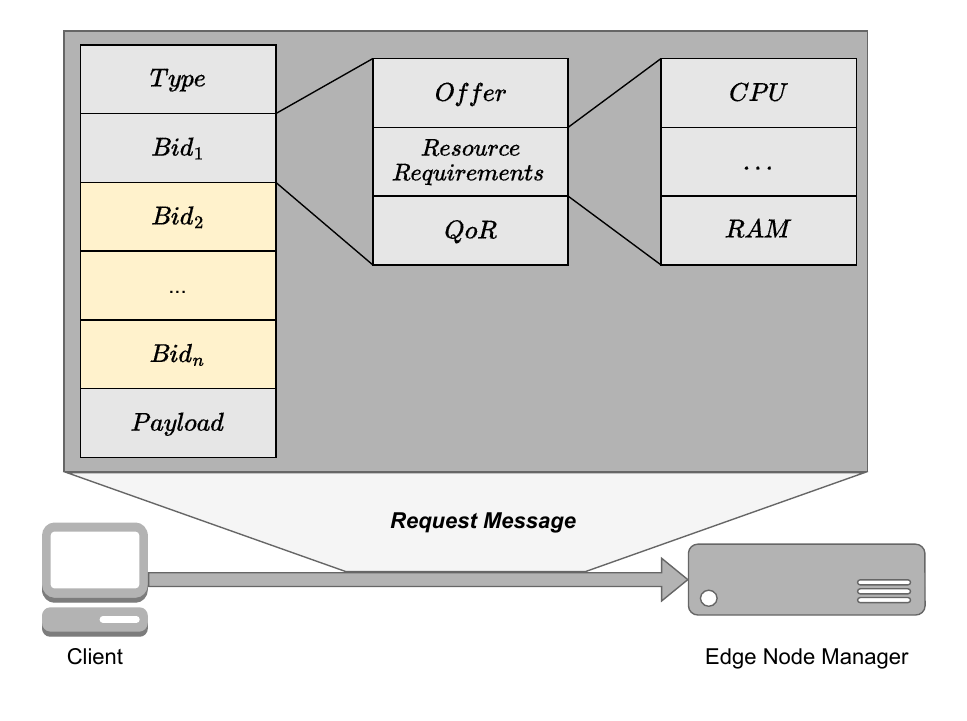}
    \caption{Request message structure between client and edge node manager. Each bid specifies a variant with its resource requirements and QoR level.}
    \label{fig:RequestMessage}
\end{figure}

Clients submit tasks through a structured request message (\Cref{fig:RequestMessage}). Each message contains a \emph{Type} identifying the task or service to be executed and a \emph{Payload} carrying the task-specific input data. In addition, the client provides a set of \emph{Bids}. Each bid refers to a concrete execution variant and contains an associated \emph{Offer} describing the variant's parameters, including resource requirements (e.g., CPU and RAM) and QoR-relevant settings. These bids encode the tenant’s \gls{WTP} for the respective variant. This information enables the provider to evaluate the request under the current system state and either select an appropriate variant or decline the request. Runtime operation is request-driven. Upon receiving a request, the manager evaluates feasible variants under the current capacity situation and forwards the selected variant to an edge node for execution. The node reserves the required resources, executes the task, and releases them upon completion, while continuously reporting utilization.
\section{Approach}\label{sec:Approach}
Building on the system architecture described in \Cref{sec:SystemDesign}, PRICE implements a QoR-aware mechanism for selecting execution variants under hard capacity constraints at the edge. For each request, tenants submit a finite set of execution variants together with bids. On the provider side, the system computes a utilization-dependent price signal and evaluates submitted bids to determine which variant, if any, to execute. If a variant is selected, the corresponding charge is returned in the response.

\subsection{Bid Submission and Request Construction}\label{sec:ApproachBids}

PRICE uses sealed (non-observable) bids at the variant level. Sealed bids prevent tenants from strategically adjusting their offers in response to observed competitor bids and avoid the multiple communication rounds that iterative bidding would require. Both properties are essential for latency-sensitive edge applications, where request handling must complete within tight time budgets. For each request, the client submits a list of variants, each described by its resource demand profile $R_{t,k}$ and a bid value $b_{t,k}$. The server then selects one of the submitted variants based on the current system state, the submitted bids, and the computed acceptance price. Tenants act non-cooperatively, each submitting bids independently without observing server-side prices or other tenants' bids, and without coordinating with other tenants. Preferences over variants are encoded as bid values derived from a tenant-specific \gls{WTP}.

Bids are derived deterministically from a tenant-specific \gls{WTP}. We model this \gls{WTP} via a function $W_t(\cdot)$ that maps a resource demand vector to a monetary bid value. The bid $b_{t,k}$ operationalizes the tenant's preference from \cref{sec:ProblemFormulation} through a resource-based \gls{WTP} function that approximates the value $V_t(q_{t,k})$ without requiring knowledge of the server-side price signal. This approximation is valid under the assumption that resource demand $R_{t,k}$ and achievable \gls{qor} $q_{t,k}$ are monotonically related for a given tenant, meaning a variant requiring more resources yields a higher result quality, so \gls{WTP} for resources serves as a proxy for $V_t(q_{t,k})$. For each variant $k$, we obtain
\[
    b_{t,k} = W_t(R_{t,k}).
\]
To capture different preference profiles, we use weighted, concave willingness-to-pay functions with weights \(w_i>0\) per resource type \(i\in I\):
\[
\begin{aligned}
W_t^\text{low}(R_{t,k})   &= \sum_{i \in I} w_i \,R_{t,k}[i]^{3/4},\\
W_t^\text{medium}(R_{t,k})&= \sum_{i \in I} w_i \,R_{t,k}[i]^{1/2},\\
W_t^\text{high}(R_{t,k})  &= \sum_{i \in I} w_i \,R_{t,k}[i]^{1/4}.
\end{aligned}
\]
The exponents control the curvature of the WTP functions and, therefore, how bid values change across variants with different resource demands. Since the functions are concave, they model diminishing marginal willingness to pay for additional resources. Different WTP levels are represented through the calibrated bid level and resource weights \(w_i\), which determine the relative importance of resource types and can be derived from market prices for comparable compute services. This allows tenants to express economically meaningful preferences without knowledge of the server-side pricing function. 

Algorithm~\ref{alg:client_cycle} shows the client-side construction of the \texttt{RequestMessage}. In addition to \emph{Type} and \emph{Payload}, the message contains a list of \emph{Bids}; each bid refers to a specific variant and conveys its resource requirements together with the corresponding bid value. Optionally, QoR-related variant parameters can be included if required for execution by the service.

\begin{algorithm2e}[t]
\DontPrintSemicolon
\caption{Client-side task submission}
\label{alg:client_cycle}

\SetKwInOut{KwIn}{Input}
\KwIn{Tenant identifier \(t\); set of permissible execution variants \(\mathcal{K}_t\) with demands \(R_{t,k}\); willingness-to-pay function \(W_t(\cdot)\); task type \texttt{Type}; task payload \texttt{Payload}.}

\SetKwFunction{Submit}{SubmitToManager}

\(B_t \gets [\,]\)\;
\ForEach{\(k \in \mathcal{K}_t\)}{
  \(b_{t,k} \gets W_t(R_{t,k})\)\tcp*{deterministic bid value}
  \(B_t \gets B_t \cup \{(k, R_{t,k}, b_{t,k})\}\)\;
}

resp \(\gets\) \Submit{\texttt{RequestMessage}\{Tenant=\(t\), Type=\texttt{Type}, Payload=\texttt{Payload}, Variants=\(B_t\)\}}\;

\uIf{resp.\texttt{status} \(=\) \texttt{SUCCESS}}{
  process result (e.g., task output)\;
}
\Else{
  handle rejection according to client policy (e.g., execute locally)\;
}
\end{algorithm2e}

\subsection{Utilization-Dependent Acceptance Price}\label{sec:ApproachPricing}
The provider evaluates execution variants relative to the current resource state of the edge node. Let \(u[i] \in [0,c_i]\) denote the current utilization of resource \(i \in I\), derived from telemetry and ongoing reservations. If a variant \(k\) of the submitted request is accepted, utilization increases by an incremental allocation \(\Delta u_{t,k}\). In our model, this incremental allocation equals the requested demand:
\[
\Delta u_{t,k}[i] = R_{t,k}[i] \quad \forall i \in I.
\]

The provider computes a utilization-dependent acceptance price $P(u, \Delta u)$ as the sum of a base term and a progressive, utilization-dependent component:
\[
P(u,\Delta u) = C_{\mathrm{base},t,k} + P_{\mathrm{util}}(u,\Delta u),
\]
where
\[
P_{\mathrm{util}}(u,\Delta u) = \sum_{i\in I}\int_{u[i]}^{u[i]+\Delta u[i]} CM'_i(c_i,x)\,dx .
\]
The term $C_{\mathrm{base},t,k}$ captures non-congestion-dependent per-request costs, such as orchestration overheads or empirically measured execution costs that depend on the selected variant. The function $CM'_i(c_i, x)$ models the marginal cost of increasing utilization of resource $i$ at utilization level $x$. The integral accumulates marginal costs from the current utilization to the post-allocation utilization, yielding a progressive price signal that penalizes allocations close to capacity. The margin score $m_{t,k} = P_{t,k} - C_{\mathrm{base},t,k}$ isolates the utilization-dependent component by subtracting the variant-specific base cost. Maximizing $m_{t,k}$ among bid-acceptable candidates therefore selects the variant that maximizes the utilization-dependent revenue contribution.

We instantiate \(CM'_i(\cdot)\) using parametric, monotonically increasing function families to capture different curvature characteristics:
\[
\begin{aligned}
\text{Linear:}      & \quad CM'_i(c_i, x) = a_i \left(\frac{x}{c_i}\right), \\
\text{Exponential:} & \quad CM'_i(c_i, x) = a_i \left(\beta^{x/c_i}-1\right), \quad \beta>1, \\
\text{Cubic:}       & \quad CM'_i(c_i, x) = a_i \left(\frac{x}{c_i}\right)^3 .
\end{aligned}
\]
Here, \(a_i>0\) and \(\beta>1\) control steepness and curvature.

To control the overall price level while preserving the curvature of the chosen function family, we introduce a provider-side \emph{unit-cost cap} vector $p_{\mathrm{cap}}[i]$ for $i \in I$. Intuitively, $p_{\mathrm{cap}}[i]$ specifies a target unit price level for resource $i$, and we use it to calibrate the steepness coefficients $a_i$ without changing the functional form of $CM'_i$. We apply $p_{\mathrm{cap}}$ as a normalization target by scaling $a_i$ such that the utilization-dependent component of allocating the entire normalized capacity of resource $i$ in a single request equals $p_{\mathrm{cap}}[i]$ under $c_i = 1$:
\[
\begin{aligned}
P_{\mathrm{util}}(u,\Delta u) &= p_{\mathrm{cap}}[i], \\
u[i] &= 0,\quad \Delta u[i]=1, \\
u[\ell] &= \Delta u[\ell] = 0 \quad \forall \ell\in I \setminus \{i\}.
\end{aligned}
\qquad \text{(Cap)}
\]

For a given request from tenant \(t\) and variant \(k\), the provider computes an acceptance price at the current utilization state \(u\). This price represents the provider-side cost signal for allocating the variant's required resources on top of the current load. Because the incremental allocation equals the requested demand, \(\Delta u_{t,k}=R_{t,k}\), we obtain
\[
P_{t,k} = P(u,\Delta u_{t,k}) = C_{\mathrm{base},t,k}+P_{\mathrm{util}}(u,R_{t,k}).
\]

\subsection{Server-Side Variant Selection}\label{sec:ApproachSelection}
The provider processes requests sequentially upon arrival and first checks feasibility under hard capacity constraints. For each resource type \(i\in I\), it considers the currently available headroom implied by capacity \(c_i\) and current utilization \(u[i]\). A variant \(k\in\mathcal{K}_t\) is feasible if
\[
u[i] + R_{t,k}[i] \le c_i \quad \forall i\in I.
\]
If no feasible variant exists, the request is rejected.

PRICE does not maintain a server-side request queue. If none of the submitted variants can be executed due to insufficient resources or missing hardware capabilities, the request is rejected immediately. 
Clients must therefore provide a fallback execution strategy, such as local execution, forwarding to another edge node, or cloud offloading, instead of waiting in an unbounded queue.
This reflects the heterogeneous nature of edge environments, where nodes may lack specialized hardware (e.g., GPU accelerators) required by some variants. By avoiding server-side queuing, PRICE allows applications to opportunistically utilize additional edge resources when available while preserving predictable response times under resource scarcity.

\begin{algorithm2e}[t]
\DontPrintSemicolon
\caption{PRICE processing of an incoming submission}
\label{alg:edge_cycle}

\SetKwInOut{KwIn}{Input}
\KwIn{capacity vector \(c\); current utilization vector \(u\); price function \(P(u,\Delta u)\); utilization-dependent component \(P_{\mathrm{util}}(u,\Delta u)\).}

\SetKw{On}{on receive}
\SetKwFunction{Respond}{Respond}
\SetKwFunction{Reserve}{ReserveResources}
\SetKwFunction{Release}{ReleaseResources}
\SetKwFunction{Execute}{ExecuteTask}

\On~\texttt{RequestMessage}\{Tenant=\(t\), Type, Payload, Variants=\(B_t\)\}\;
\BlankLine

\textbf{Feasibility:}\;
\(\mathcal{F} \gets \{\, k \mid (k,R_{t,k},b_{t,k})\in B_t \ \wedge\  u[i] + R_{t,k}[i] \le c_i\ \forall i\in I \,\}\)\;
\If{\(\mathcal{F} = \emptyset\)}{
  \Respond{status=\texttt{FAILURE}, reason=\texttt{insufficient\_capacity}}\;
  \Return\;
}

\textbf{Bid-acceptability and selection:}\;
\(k^\star \gets \bot,\; P^\star \gets 0,\; m^\star \gets -\infty\)\;
\ForEach{\(k \in \mathcal{F}\)}{
  \(P_{t,k} \gets P(u, R_{t,k})\)\;
  \If{\(b_{t,k} \ge P_{t,k}\)}{
    \(m_{t,k} \gets P_{\mathrm{util}}(u,R_{t,k})\)\;
    \If{\(m_{t,k} > m^\star\)}{
      \(m^\star \gets m_{t,k},\; k^\star \gets k,\; P^\star \gets P_{t,k}\)\;
    }
  }
}

\If{\(k^\star = \bot\)}{
  \Respond{status=\texttt{FAILURE}, reason=\texttt{bid\_below\_price}}\;
  \Return\;
}

\textbf{Execution:}\;
\Reserve{\(R_{t,k^\star}\)}\;
result \(\gets\) \Execute{Type, Payload, variant=\(k^\star\)}\;
\Release{\(R_{t,k^\star}\)}\;
\Respond{status=\texttt{SUCCESS}, variant=\(k^\star\), price=\(P^\star\), result=result}\;
\end{algorithm2e}

For each feasible variant, the provider then computes the acceptance price \(P_{t,k}\). A feasible variant is bid-acceptable if the bid covers the acceptance price:
\[
b_{t,k} \ge P_{t,k}.
\]

Among all feasible and bid-acceptable variants, the provider accepts at most one variant. PRICE ranks candidates by the utilization-dependent margin
\[
m_{t,k}=P_{\mathrm{util}}(u,R_{t,k}),
\]
which is equivalent to \(m_{t,k}=P_{t,k}-C_{\mathrm{base},t,k}\), and selects the candidate with maximal \(m_{t,k}\). If no bid-acceptable variant exists, the request is rejected. This selection maximizes the utilization-dependent revenue contribution among acceptable candidates.

Algorithm~\ref{alg:edge_cycle} implements this pipeline. We use \(\bot\) as a sentinel value for \(k^\star\) to indicate that no feasible and bid-acceptable variant was selected; in this case, the request is rejected. Upon acceptance, the selected edge node reserves the resources required by the chosen variant, executes the task, releases the resources upon completion, and returns a \texttt{BidResponse} reporting the selected variant, the charged value $P_{t,k^\star}$, and the task result.

\subsection{Telemetry Feedback and Self-Regulating Behavior}\label{sec:ApproachFeedback}
The acceptance price is state-dependent and is continuously updated based on the latest telemetry data. Edge nodes periodically report resource utilization, allowing the provider to keep the utilization vector $u$ up to date. As utilization increases, marginal costs $CM'_i$ and thus the acceptance price $P(\cdot)$ increase, making resource-intensive variants less likely to be bid-acceptable. As utilization decreases, the acceptance price drops, and higher-quality variants become executable again. This self-regulating behavior drives the system toward sustained high utilization near capacity without requiring explicit coordination or prior knowledge of the workload. In this way, PRICE aligns non-cooperative tenant bidding with the provider's objective of avoiding persistent overload while maintaining high resource utilization. The empirical stability of this behavior is demonstrated in \Cref{sec:Evaluation}.

\begin{table*}[!b]
\centering
\small
\caption{Task types and resource profiles per discrete QoR/variant level \(k\in\{1,\dots,5\}\) as configured in the client. The default willingness-to-pay is defined for the default variant \(k=5\).}
\label{tab:task_profiles}
\begin{tabular}{l c c c }
\hline
Application & Default WTP [\$/h] & Default RAM [GB] & CPU cores per \(k=1..5\)\\
\hline
Object detection  & 6.12 & 2.02 & \(\{0.554,\,0.894,\,1.600,\,2.890,\,3.958\}\) \\
Sensor fusion     & 4.59 & 1.52 & \(\{0.500,\,1.000,\,1.800,\,2.500,\,3.200\}\) \\
NLP preprocessing & 3.06 & 1.02 & \(\{0.300,\,0.600,\,1.000,\,1.500,\,2.000\}\) \\
Anomaly detection & 3.82 & 0.82 & \(\{0.800,\,1.500,\,2.500,\,3.500,\,4.500\}\) \\
\hline
\end{tabular}
\end{table*}

\section{Evaluation}
\label{sec:Evaluation}
We evaluate PRICE under sustained overload, where aggregate demand continuously exceeds node capacity. This setting reflects the operational scenario that PRICE targets, namely an edge node that must operate close to its resource limits while serving multiple independent tenants with heterogeneous workloads. Under these conditions, no mechanism can accept all incoming requests, and the relevant question is how efficiently available capacity is converted into completed work and how predictably the system behaves under load. The evaluation analyzes whether pricing-based QoR incentives improve resource utilization, system stability, and adaptive QoR selection compared to static and QoR-unaware baselines.

The evaluation is structured around the following research questions.

\textbf{RQ1: Throughput and Rejection under Overload.}
Under sustained overload, how do the accepted throughput and rejection behavior of PRICE compare to the baselines, and how are these outcomes affected by the provider's unit-cost cap and by the composition of the client population in terms of WTP?

\textbf{RQ2: Stability and Resource Utilization.}
Does PRICE ensure stable operation under overload, reflected in persistently high but controlled CPU utilization without unstable oscillations or collapse?

\textbf{RQ3: QoR Adaptation and Monetization.}
How does PRICE adapt delivered QoR under contention, and how does this adaptation affect provider revenue across client populations with different WTP compositions?

Taken together, these research questions examine whether PRICE under overload efficiently converts scarce CPU resources into completed work, maintains stable utilization, and uses QoR adaptation as a control dimension under heterogeneous client WTP profiles.

\begin{figure*}[t]
\centering
\includegraphics[width=.9\textwidth]{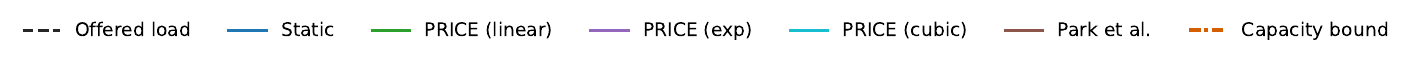}\par
\begin{tabular}{@{}lccc@{}}
\toprule
& \small\textbf{CAP=1.0} & \small\textbf{CAP=1.5} & \small\textbf{CAP=2.0} \\
\midrule
\multirow{2}{*}{\raisebox{2.8cm}{\rotatebox{90}{\textbf{const}}}} &
\includegraphics[width=.28\textwidth]{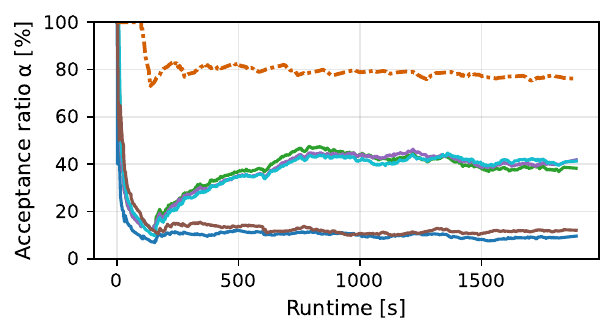} &
\includegraphics[width=.28\textwidth]{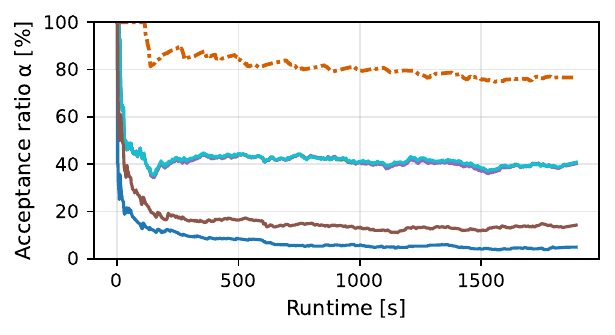} &
\includegraphics[width=.28\textwidth]{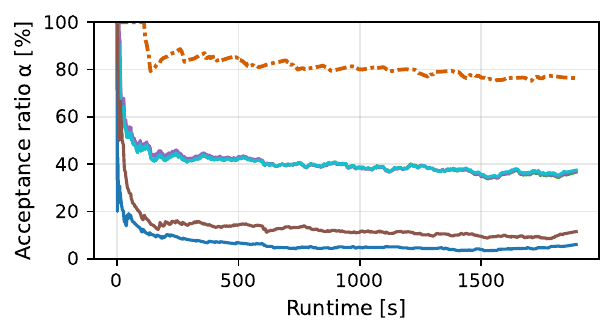} \\
&
\includegraphics[width=.28\textwidth]{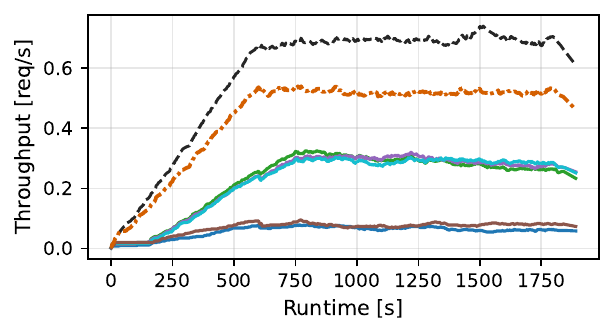} &
\includegraphics[width=.28\textwidth]{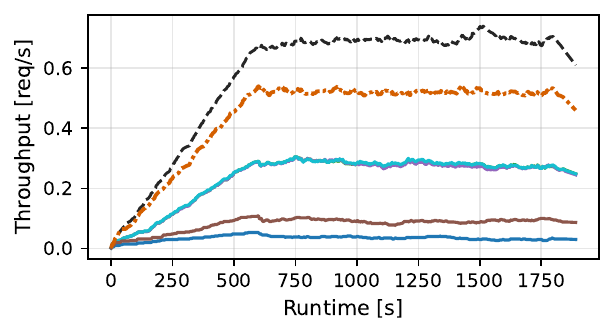} &
\includegraphics[width=.28\textwidth]{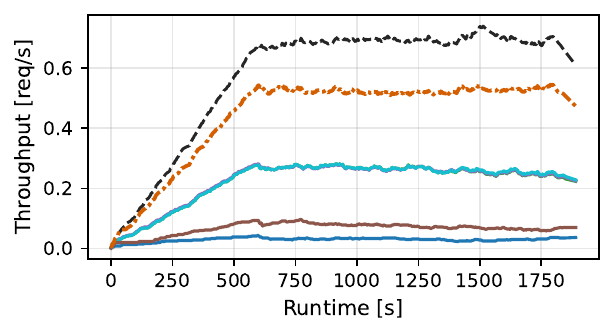} \\
\midrule
\multirow{2}{*}{\raisebox{2.2cm}{\rotatebox{90}{\textbf{Pareto}}}} &
\includegraphics[width=.28\textwidth]{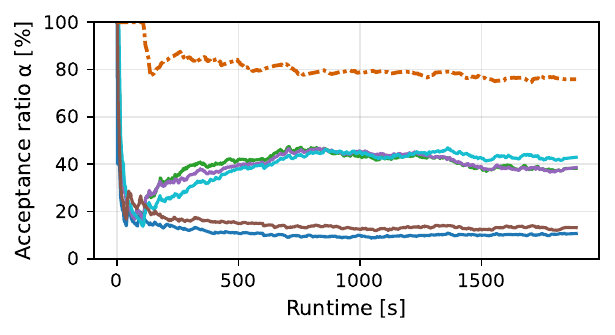} &
\includegraphics[width=.28\textwidth]{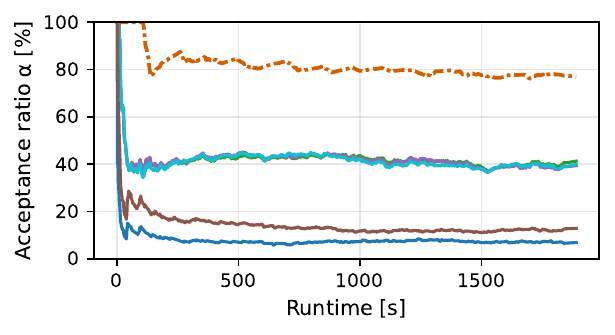} &
\includegraphics[width=.28\textwidth]{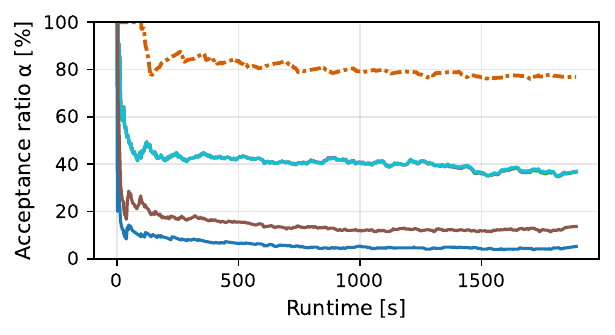} \\
&
\includegraphics[width=.28\textwidth]{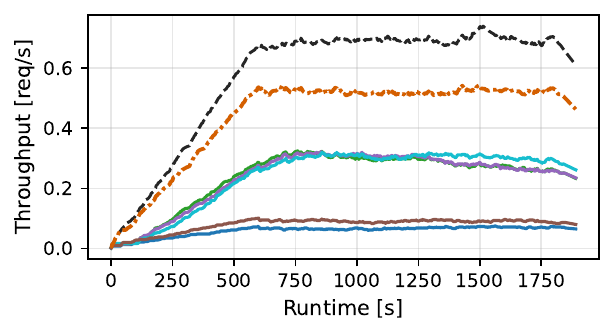} &
\includegraphics[width=.28\textwidth]{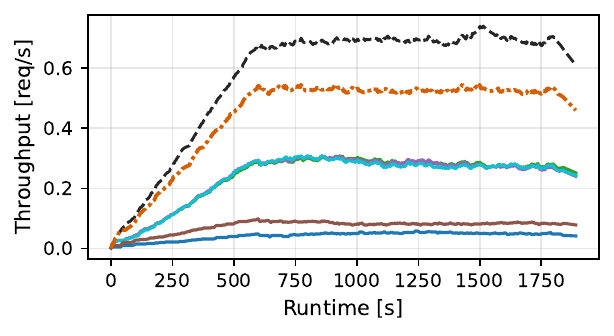} &
\includegraphics[width=.28\textwidth]{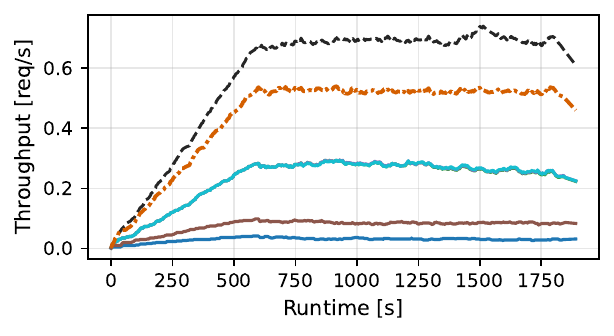} \\
\bottomrule
\end{tabular}
\caption{Acceptance ratio $\alpha$ (top row per distribution) and accepted throughput $\lambda_{\mathrm{acc}}$ (bottom row) under sustained overload, mean task duration $d=60\,\mathrm{s}$. Columns vary the server price cap~(CAP). Sliding window: $w=600\,\mathrm{s}$.}
\label{fig:rq1_d60_grid}
\end{figure*}

\subsection{Test Setup}

Our experimental setup consists of three servers acting as edge nodes, each with 24 CPU cores and 64\,GB RAM.
Services run as Docker containers for process isolation.
Contention is, by construction, CPU-driven because variants primarily differ in CPU demand; RAM and hardware capabilities still constrain admission, and requests are rejected if no variant fits all resource and capability constraints.

We vary the provider-side unit-cost cap $p_{\mathrm{cap}}[\mathrm{cpu}] \in \{1.0, 1.5, 2.0\}$, which scales the utilization-dependent pricing function as described in \cref{sec:ApproachPricing}, while preserving the curvature across the three pricing families. For brevity, we refer to this value as $\mathrm{CAP}$ in the following.

Clients generate tasks from four application types, each with five discrete execution variants $k \in \{1, \dots, 5\}$, where larger $k$ corresponds to higher QoR and resource consumption. These discrete levels abstract over the normalized QoR scale introduced in Section~\ref{sec:Background}, with $k=1$ corresponding to the lowest acceptable quality and $k=5$ to the highest available variant. Resource profiles and default \gls{WTP} values are summarized in \cref{tab:task_profiles}. The default WTP values are calibrated from real market prices for comparable cloud inference services, following the methodology described in \Cref{sec:Approach}. WTP values for all application types are scaled proportionally according to their default resource demand and task duration, as listed in \cref{tab:task_profiles}.

Task arrivals follow a Poisson process with exponentially distributed inter-arrival times. Runtime behavior is modeled separately using task durations that are constant (\emph{const}), exponentially distributed (\emph{exp}), or Pareto-distributed (\emph{Pareto}). These durations parameterize submitted application tasks, not the arrival process, and are not used by the server for pricing or feasibility decisions. The server reserves resources according to the selected variant and releases them after execution, while the observed container runtime is logged separately. We evaluate mean durations $d \in \{15,30,60\}\,\mathrm{s}$ with corresponding arrival rates and four WTP mixtures $(n_h,n_m,n_l) \in \{(4,4,4),(6,3,3),(3,6,3),(3,3,6)\}$.

We compare Static, Park et al.~\cite{park_real-time_2024}, and three PRICE variants with linear, cubic, and exponential pricing functions. Static always executes the highest-quality variant regardless of load. Park et al.~\cite{park_real-time_2024} introduces utilization-dependent pricing, but treats each request as a fixed-configuration submission, leaving binary accept-or-reject as the only control lever under high load. PRICE introduces result-quality flexibility as an orthogonal control dimension, allowing the pricing signal to shape not only whether a request is served but at which quality level. Per-task outcomes and CPU utilization are logged; all data are publicly available at~\cite{gropengieser_artifacts_2026}.

\begin{table*}[t]
\centering
\setlength{\tabcolsep}{3pt}
\resizebox{\linewidth}{!}{%
\begin{tabular}{@{}lcccccccccccccccccc@{}}
\toprule
\textbf{Method} & \multicolumn{6}{c}{\textbf{Duration: const}} & \multicolumn{6}{c}{\textbf{Duration: exp}} & \multicolumn{6}{c}{\textbf{Duration: pareto}} \\
\cmidrule(lr){2-7} \cmidrule(lr){8-13} \cmidrule(lr){14-19}
 & \multicolumn{2}{c}{\textbf{CAP=1.0}} & \multicolumn{2}{c}{\textbf{CAP=1.5}} & \multicolumn{2}{c}{\textbf{CAP=2.0}} & \multicolumn{2}{c}{\textbf{CAP=1.0}} & \multicolumn{2}{c}{\textbf{CAP=1.5}} & \multicolumn{2}{c}{\textbf{CAP=2.0}} & \multicolumn{2}{c}{\textbf{CAP=1.0}} & \multicolumn{2}{c}{\textbf{CAP=1.5}} & \multicolumn{2}{c}{\textbf{CAP=2.0}} \\
\cmidrule(lr){2-3} \cmidrule(lr){4-5} \cmidrule(lr){6-7} \cmidrule(lr){8-9} \cmidrule(lr){10-11} \cmidrule(lr){12-13} \cmidrule(lr){14-15} \cmidrule(lr){16-17} \cmidrule(lr){18-19}
 & $\alpha$ [\%] & $\lambda_{{\mathrm{{acc}}}}$ [req/s] & $\alpha$ [\%] & $\lambda_{{\mathrm{{acc}}}}$ [req/s] & $\alpha$ [\%] & $\lambda_{{\mathrm{{acc}}}}$ [req/s] & $\alpha$ [\%] & $\lambda_{{\mathrm{{acc}}}}$ [req/s] & $\alpha$ [\%] & $\lambda_{{\mathrm{{acc}}}}$ [req/s] & $\alpha$ [\%] & $\lambda_{{\mathrm{{acc}}}}$ [req/s] & $\alpha$ [\%] & $\lambda_{{\mathrm{{acc}}}}$ [req/s] & $\alpha$ [\%] & $\lambda_{{\mathrm{{acc}}}}$ [req/s] & $\alpha$ [\%] & $\lambda_{{\mathrm{{acc}}}}$ [req/s] \\
\midrule
\multicolumn{19}{@{}l}{\textit{low-WTP heavy (3,3,6)}}\\
\quad Static & 8.8 & 0.061 & 6.4 & 0.045 & 4.2 & 0.030 & 8.2 & 0.057 & 4.9 & 0.034 & 4.8 & 0.033 & 10.2 & 0.071 & 6.0 & 0.042 & 4.6 & 0.032 \\
\quad PRICE (linear) & 40.1 & 0.280 & \textbf{38.3} & \textbf{0.267} & \textbf{32.4} & \textbf{0.227} & \textbf{42.6} & \textbf{0.298} & 37.8 & 0.264 & \textbf{32.5} & \textbf{0.227} & \textbf{44.4} & \textbf{0.310} & 39.8 & 0.278 & \textbf{34.1} & \textbf{0.238} \\
\quad PRICE (exp) & \textbf{40.9} & \textbf{0.286} & 37.3 & 0.260 & \textbf{32.4} & 0.226 & 40.4 & 0.282 & \textbf{38.1} & \textbf{0.266} & 32.3 & 0.226 & 44.3 & \textbf{0.310} & \textbf{40.8} & \textbf{0.285} & 33.9 & 0.237 \\
\quad PRICE (cubic) & 40.7 & 0.284 & 37.1 & 0.259 & \textbf{32.4} & 0.226 & 38.8 & 0.271 & 37.2 & 0.260 & 32.0 & 0.224 & 43.3 & 0.303 & 40.5 & 0.283 & 33.7 & 0.235 \\
\quad Park et al. & 13.4 & 0.094 & 11.7 & 0.082 & 13.3 & 0.093 & 12.9 & 0.090 & 11.3 & 0.079 & 12.0 & 0.084 & 10.9 & 0.076 & 9.4 & 0.066 & 12.2 & 0.086 \\
\addlinespace[3pt]
\cmidrule{1-19}
\addlinespace[3pt]
\multicolumn{19}{@{}l}{\textit{mid-WTP heavy (3,6,3)}}\\
\quad Static & 8.9 & 0.062 & 5.5 & 0.038 & 4.4 & 0.031 & 9.4 & 0.066 & 6.5 & 0.045 & 4.0 & 0.028 & 10.5 & 0.073 & 7.0 & 0.049 & 4.6 & 0.032 \\
\quad PRICE (linear) & \textbf{39.3} & \textbf{0.274} & 40.2 & \textbf{0.281} & 36.9 & 0.258 & 37.9 & 0.265 & \textbf{40.2} & \textbf{0.281} & 36.9 & 0.258 & 39.3 & 0.274 & \textbf{41.4} & \textbf{0.289} & 37.7 & 0.263 \\
\quad PRICE (exp) & 38.3 & 0.267 & 39.2 & 0.274 & 37.2 & 0.260 & 38.5 & 0.269 & 40.0 & 0.279 & \textbf{37.2} & \textbf{0.260} & \textbf{41.1} & \textbf{0.287} & 40.8 & 0.285 & \textbf{37.9} & \textbf{0.265} \\
\quad PRICE (cubic) & 38.5 & 0.269 & \textbf{40.3} & \textbf{0.281} & \textbf{37.4} & \textbf{0.261} & \textbf{39.9} & \textbf{0.279} & 39.2 & 0.274 & \textbf{37.2} & \textbf{0.260} & 39.9 & 0.279 & 41.0 & 0.287 & 37.8 & 0.264 \\
\quad Park et al. & 13.8 & 0.096 & 11.0 & 0.077 & 12.1 & 0.085 & 11.3 & 0.079 & 14.9 & 0.104 & 11.4 & 0.080 & 13.2 & 0.092 & 10.1 & 0.070 & 10.5 & 0.073 \\
\addlinespace[3pt]
\cmidrule{1-19}
\addlinespace[3pt]
\multicolumn{19}{@{}l}{\textit{balanced (4,4,4)}}\\
\quad Static & 10.0 & 0.070 & 5.7 & 0.040 & 4.9 & 0.034 & 8.7 & 0.061 & 5.7 & 0.040 & 3.6 & 0.025 & 9.7 & 0.068 & 7.1 & 0.050 & 4.6 & 0.032 \\
\quad PRICE (linear) & 39.5 & 0.276 & 37.8 & 0.264 & 34.9 & 0.244 & 40.5 & 0.283 & \textbf{38.7} & \textbf{0.270} & \textbf{35.2} & \textbf{0.246} & \textbf{43.3} & \textbf{0.303} & 37.9 & 0.265 & \textbf{36.0} & \textbf{0.251} \\
\quad PRICE (exp) & \textbf{40.6} & \textbf{0.284} & \textbf{38.2} & \textbf{0.267} & 35.1 & 0.245 & \textbf{40.6} & \textbf{0.284} & 37.9 & 0.264 & 34.9 & 0.244 & 41.9 & 0.292 & 38.6 & 0.270 & 35.8 & 0.250 \\
\quad PRICE (cubic) & 37.9 & 0.265 & 37.8 & 0.264 & \textbf{35.5} & \textbf{0.248} & 39.5 & 0.276 & 38.4 & 0.268 & \textbf{35.2} & \textbf{0.246} & 40.9 & 0.286 & \textbf{40.1} & \textbf{0.280} & \textbf{35.9} & \textbf{0.251} \\
\quad Park et al. & 14.0 & 0.097 & 12.8 & 0.090 & 12.6 & 0.088 & 11.6 & 0.081 & 12.1 & 0.084 & 12.7 & 0.089 & 12.3 & 0.086 & 13.7 & 0.096 & 12.4 & 0.087 \\
\addlinespace[3pt]
\cmidrule{1-19}
\addlinespace[3pt]
\multicolumn{19}{@{}l}{\textit{high-WTP heavy (6,3,3)}}\\
\quad Static & 9.2 & 0.065 & 4.8 & 0.034 & 4.3 & 0.030 & 10.2 & 0.071 & 5.3 & 0.037 & 4.6 & 0.032 & 9.9 & 0.069 & 7.3 & 0.051 & 4.5 & 0.031 \\
\quad PRICE (linear) & 40.7 & 0.284 & 39.9 & 0.279 & 37.1 & 0.259 & 39.1 & 0.273 & 39.5 & 0.276 & \textbf{37.3} & \textbf{0.260} & 40.7 & 0.285 & \textbf{40.7} & \textbf{0.285} & 38.5 & 0.269 \\
\quad PRICE (exp) & \textbf{42.0} & \textbf{0.293} & 39.5 & 0.276 & 37.2 & 0.259 & 40.9 & 0.285 & 38.8 & 0.271 & \textbf{37.3} & \textbf{0.260} & 41.3 & 0.288 & 40.2 & 0.281 & 38.7 & \textbf{0.271} \\
\quad PRICE (cubic) & 41.6 & 0.290 & \textbf{40.3} & \textbf{0.281} & \textbf{37.4} & \textbf{0.261} & \textbf{42.3} & \textbf{0.296} & \textbf{39.8} & \textbf{0.278} & 37.1 & 0.259 & \textbf{43.7} & \textbf{0.305} & 39.6 & 0.277 & \textbf{38.8} & \textbf{0.271} \\
\quad Park et al. & 11.2 & 0.078 & 12.9 & 0.090 & 10.2 & 0.071 & 12.7 & 0.089 & 12.1 & 0.084 & 11.9 & 0.083 & 13.0 & 0.091 & 11.8 & 0.082 & 12.1 & 0.085 \\
\bottomrule
\end{tabular}}
\caption{Steady-state acceptance ratio $\alpha$ [\%] and accepted throughput $\lambda_{\mathrm{acc}}$ [req/s] under sustained overload. Duration distributions: const, exp, and pareto; mean task duration $d=60\,\mathrm{s}$, CAP $\in\{1.0/1.5/2.0\}$, sliding window $w=600\,\mathrm{s}$. Best $\alpha$ and $\lambda_{\mathrm{acc}}$ per CAP in bold.}
\label{tab:rq1_big_merged_d60}
\end{table*}

\subsection{RQ1: Throughput and Rejection under Overload}
\label{sec:rq1}

RQ1 examines system behavior under sustained overload by measuring the incoming request rate $\lambda_{\mathrm{in}}$, the accepted throughput $\lambda_{\mathrm{acc}}$, and the acceptance ratio $\alpha = \lambda_{\mathrm{acc}}/\lambda_{\mathrm{in}}$. Rates are computed over a sliding window of $w = 600\,\mathrm{s}$. The experiments intentionally operate above the edge node's capacity limits to make overload handling observable. As a result, a large gap between $\lambda_{\mathrm{in}}$ and $\lambda_{\mathrm{acc}}$ is expected, because any stable mechanism must reject requests once CPU and memory approach capacity rather than accumulating an unbounded backlog.

Figure~\ref{fig:rq1_d60_grid} shows the time evolution of $\alpha$ and $\lambda_{\mathrm{acc}}$ for mean task duration $d = 60\,\mathrm{s}$, covering both constant and Pareto-distributed runtimes across three CAP values. The initial ramp-up phase visible in all curves reflects the sliding window filling up before steady state is reached. The offered load remains stable across methods, confirming comparable demand. Differences, therefore, stem from the mechanisms' rejection behavior and the resulting conversion of offered load into accepted executions.

\begin{figure*}[b]
\centering
\includegraphics[width=0.7\textwidth]{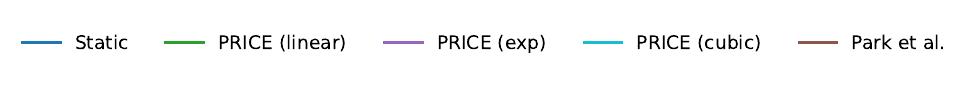}\par

\begin{tabular}{@{}ccc@{}}
\small\textbf{Duration: const} &
\small\textbf{Duration: exp} &
\small\textbf{Duration: Pareto} \\[0.2em]
\includegraphics[width=.31\textwidth]{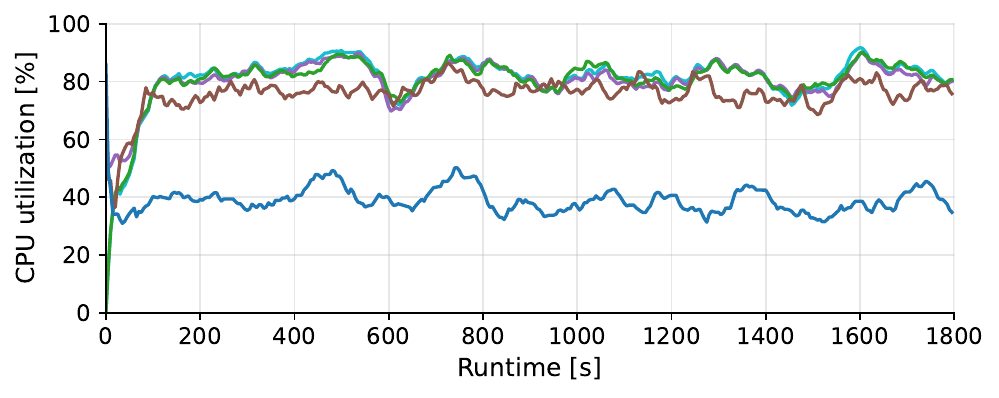} &
\includegraphics[width=.31\textwidth]{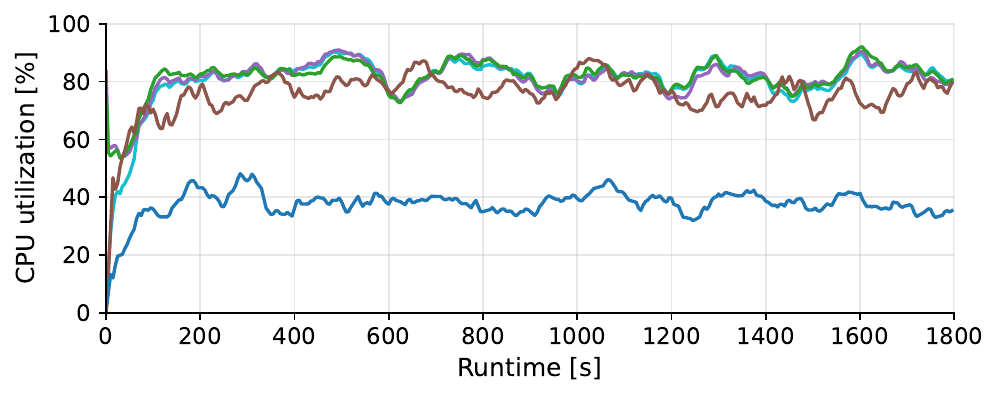} &
\includegraphics[width=.31\textwidth]{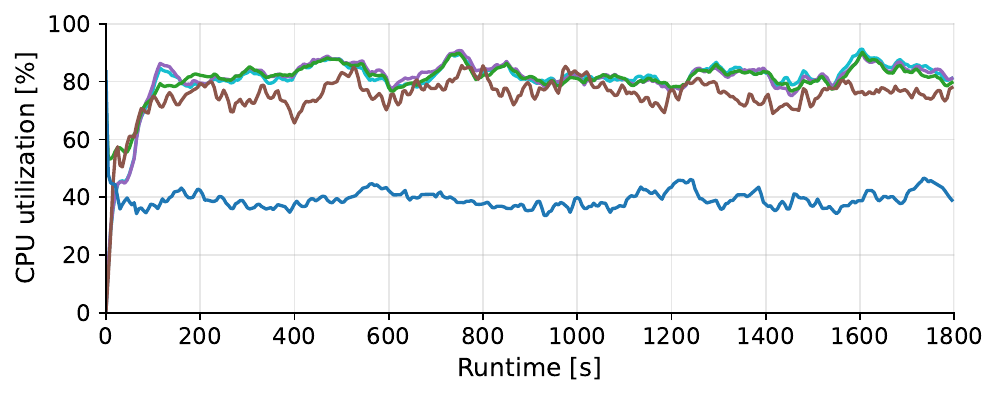} \\
\end{tabular}

\caption{CPU utilization over time under sustained overload, mean task duration $d=30\,\mathrm{s}$, $\mathrm{CAP}=1.5$. PRICE stabilizes utilization at a high controlled level across all three duration distributions, while Static remains well below capacity due to its high rejection rate. Park et al. achieves comparable utilization but at a significantly lower acceptance ratio, as shown in Table~\ref{tab:rq1_big_merged_d60}.}
\label{fig:rq2_cpu}
\end{figure*}

Across all settings, PRICE consistently outperforms both baselines. Under constant durations and $\mathrm{CAP} = 1.5$, all three PRICE variants reach a steady-state acceptance ratio of approximately $38$--$40\,\%$, compared to $5$--$10\,\%$ for Static and $11$--$13\,\%$ for Park et al. The throughput gap is similar. Under the same CAP setting, PRICE achieves \(\lambda_{\mathrm{acc}} \approx 0.26\)--\(0.28\,\mathrm{req/s}\), compared to \(0.03\)--\(0.05\,\mathrm{req/s}\) for Static. This advantage arises because PRICE admits lower-QoR variants when utilization is high, whereas Static insists on a fixed resource footprint and is therefore forced to reject most requests when the node approaches capacity. Park et al.\ also compute a utilization-dependent price, but treat each request as a fixed-configuration submission. When the acceptance price rises under high load, a request is either accepted at its fixed resource footprint or rejected outright, with no possibility of switching to a lighter variant. This leaves Park et al.\ with the same binary control lever as Static, explaining why its acceptance rate remains well below that of PRICE.

The three PRICE pricing functions---linear, exponential, and cubic---perform comparably in terms of \(\alpha\) and \(\lambda_{\mathrm{acc}}\). Although individual configurations show small differences of a few percentage points, the ranking against both baselines remains unchanged. This indicates that the dominant factor is the presence of QoR-aware variant selection itself rather than the specific curvature of the pricing curve.

The provider-side unit-cost cap CAP has a clear and systematic effect. Raising CAP from $1.0$ to $2.0$ increases the dynamic range of the pricing signal, which makes the mechanism more selective with respect to submitted bids. Under constant durations, PRICE's acceptance ratio shifts from roughly $40\,\%$ at $\mathrm{CAP} = 1.0$ to $32$--$37\,\%$ at $\mathrm{CAP} = 2.0$, indicating that a higher cap makes the acceptance price more selective and rejects a larger share of low-bid requests.
This behavior is consistent across all WTP mixture configurations shown in Table~\ref{tab:rq1_big_merged_d60}, where the ranking of mechanisms is preserved regardless of whether the population is dominated by high-, medium-, or low-WTP tenants.

Comparing constant and Pareto-distributed task durations shows that heavy-tailed runtime variability changes contention dynamics through long-running tasks. Nevertheless, PRICE maintains a substantial throughput advantage over both baselines across all CAP values and WTP compositions, confirming that the mechanism is robust to heavy-tailed service-time distributions and heterogeneous client populations.

\subsection{RQ2: Stability and Resource Utilization}

RQ2 examines whether PRICE maintains stable CPU utilization near capacity under sustained overload, and how this utilization relates to the acceptance behavior observed in RQ1.

\begin{figure*}[t]
    \centering
    \includegraphics[width=1.2\columnwidth]{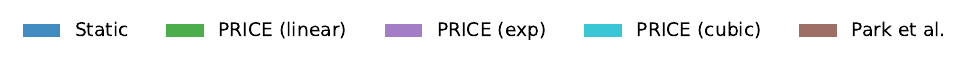}\par
    \includegraphics[width=.95\linewidth]{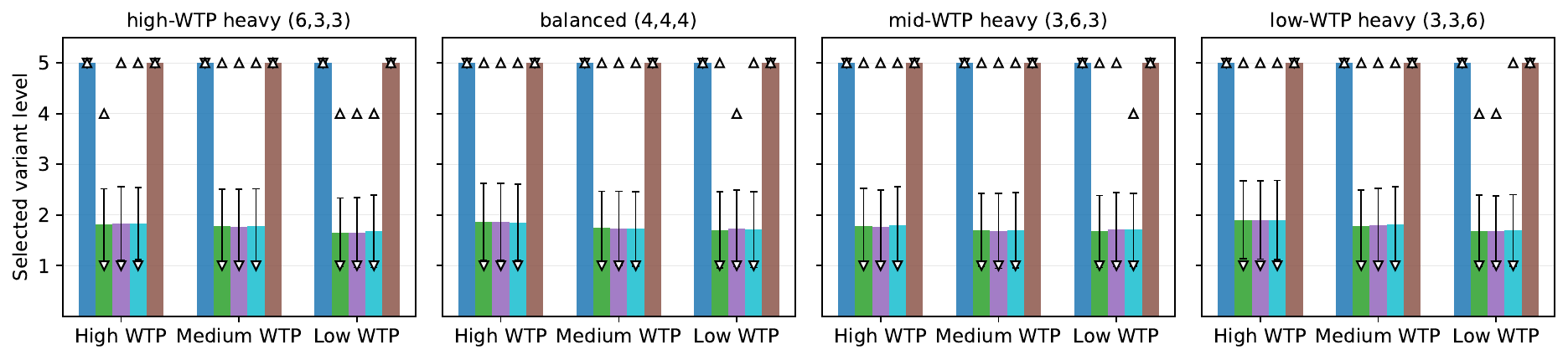}
    \caption{Mean selected variant level per WTP group and population mixture under sustained overload ($d=30\,\mathrm{s}$, all CAP values and duration distributions aggregated). Bars show the mean, error bars indicate one standard deviation, markers ($\triangledown$/$\triangle$) denote observed minimum and maximum.}
    \label{fig:RQ3_QoR}
\end{figure*}

Figure~\ref{fig:rq2_cpu} shows CPU utilization over time for mean task duration $d = 30\,\mathrm{s}$, which exposes the feedback dynamics more clearly than longer durations because resources are released more frequently, allowing the pricing signal to react within a shorter observation window. Across all three distributions, all three PRICE variants stabilize CPU utilization in the range of $70$--$85\,\%$ after an initial ramp-up phase of roughly $200\,\mathrm{s}$, and maintain this level without oscillations or sudden drops throughout the remaining observation window. Static, by contrast, settles near $35$--$45\,\%$, which at first glance may appear as controlled behavior but must be interpreted in conjunction with the acceptance ratios from RQ1. Static achieves this utilization level not through efficient scheduling but through the inverse. It accepts a small fraction of incoming requests at full resource demand, leaving the majority of capacity idle. Park et al. reaches utilization levels comparable to PRICE under some configurations, yet Table~\ref{tab:rq1_big_merged_d60} shows that this comes with an acceptance ratio of only $11$--$15\,\%$, whereas PRICE simultaneously achieves $38$--$42\,\%$. The utilization figures of Park et al. are therefore driven by the same effect as Static: few accepted tasks with large, fixed resource footprints occupy the node, while the majority of demand is rejected. PRICE, in contrast, fills available capacity by accepting many tasks whose resource footprints are continuously shaped by the pricing signal, allowing lower-QoR variants under high load to sustain throughput without exceeding capacity.

The feedback loop between utilization and acceptance price is visible in the ramp-up behavior. As the node fills during the first $200\,\mathrm{s}$, the acceptance price rises, and variant selection shifts toward lighter variants, allowing the system to continue accepting requests rather than rejecting them. Once steady state is reached, the pricing signal holds the operating point near capacity without manual tuning or explicit coordination. This behavior is consistent across all three duration distributions, confirming that the feedback mechanism is robust to variance introduced by Pareto-distributed task durations.

\subsection{RQ3: QoR Adaptation and Monetization}
RQ3 examines whether PRICE adapts result quality under contention and whether this adaptation translates into measurable provider revenue.

\Cref{fig:RQ3_QoR} shows the mean selected variant level per WTP group, aggregated over all CAP values and duration distributions with $d=30\,\text{s}$. Note that the figure covers only accepted requests. Rejected requests, which occur more frequently for low-WTP clients under high load, are not reflected in the variant-level distributions. Across all population compositions, the mean selected variant level is similar across WTP groups for all PRICE variants, indicating that the pricing mechanism does not impose a strict quality cutoff based on WTP alone. 
Static and Park et al.\ fix execution at the highest variant regardless of WTP, because neither mechanism ties resource selection to client-side WTP signals.
The QoR adaptation visible in \Cref{fig:RQ3_QoR} indicates WTP-based service differentiation rather than fairness-guaranteed allocation; it is reported at WTP-group level and is not engineered through explicit quotas or fairness constraints.
While this paper uses discrete execution variants, the same principle applies to continuous quality ranges as demonstrated in systems such as MARQ~\cite{gropengieser_marq_2025}, where QoR is expressed as a normalized percentage.

The \gls{WTP} values used in the evaluation, summarized in \Cref{tab:task_profiles}, are expressed in dollars per hour and calibrated from real market prices for comparable inference services. Bid values per request are derived from these hourly rates via \(W_t(\cdot)\), as described in \Cref{sec:Approach}. Revenue totals and rates should be interpreted relative to mechanisms rather than as absolute monetary values.

\begin{table*}[t]
\centering
\setlength{\tabcolsep}{4pt}
\resizebox{\linewidth}{!}{%
\begin{tabular}{@{}lcccccccccccccccccc@{}}
\toprule
\textbf{Method} & \multicolumn{6}{c}{\textbf{Duration: const}} & \multicolumn{6}{c}{\textbf{Duration: exp}} & \multicolumn{6}{c}{\textbf{Duration: pareto}} \\
\cmidrule(lr){2-7} \cmidrule(lr){8-13} \cmidrule(lr){14-19}
 & \multicolumn{2}{c}{\textbf{CAP=1.0}} & \multicolumn{2}{c}{\textbf{CAP=1.5}} & \multicolumn{2}{c}{\textbf{CAP=2.0}} & \multicolumn{2}{c}{\textbf{CAP=1.0}} & \multicolumn{2}{c}{\textbf{CAP=1.5}} & \multicolumn{2}{c}{\textbf{CAP=2.0}} & \multicolumn{2}{c}{\textbf{CAP=1.0}} & \multicolumn{2}{c}{\textbf{CAP=1.5}} & \multicolumn{2}{c}{\textbf{CAP=2.0}} \\
\cmidrule(lr){2-3} \cmidrule(lr){4-5} \cmidrule(lr){6-7} \cmidrule(lr){8-9} \cmidrule(lr){10-11} \cmidrule(lr){12-13} \cmidrule(lr){14-15} \cmidrule(lr){16-17} \cmidrule(lr){18-19}
 & rate [\$/s] & total [\$] & rate [\$/s] & total [\$] & rate [\$/s] & total [\$] & rate [\$/s] & total [\$] & rate [\$/s] & total [\$] & rate [\$/s] & total [\$] & rate [\$/s] & total [\$] & rate [\$/s] & total [\$] & rate [\$/s] & total [\$] \\
\midrule
\multicolumn{19}{@{}l}{\textit{low-WTP heavy (3,3,6)}} \\\\
\quad Static & 0.0033 & 9.19 & 0.0019 & 5.04 & 0.0012 & 3.38 & 0.0033 & 9.03 & 0.0019 & 5.13 & 0.0011 & 3.13 & 0.0033 & 9.14 & 0.0020 & 5.42 & 0.0011 & 3.02 \\
\quad PRICE (linear) & \textbf{0.0063} & \textbf{16.83} & 0.0062 & 17.87 & 0.0054 & 15.56 & 0.0061 & 16.99 & 0.0062 & \textbf{18.08} & 0.0051 & 15.37 & 0.0058 & 15.70 & 0.0068 & 17.65 & 0.0056 & 15.39 \\
\quad PRICE (exp) & 0.0060 & 16.47 & 0.0064 & 17.84 & 0.0053 & 15.67 & \textbf{0.0061} & \textbf{17.13} & \textbf{0.0063} & 17.85 & 0.0054 & 15.71 & 0.0056 & 15.98 & 0.0068 & 17.80 & 0.0055 & 15.17 \\
\quad PRICE (cubic) & 0.0057 & 15.41 & \textbf{0.0064} & \textbf{17.92} & 0.0055 & 15.57 & 0.0059 & 16.06 & 0.0062 & 17.86 & 0.0053 & 15.23 & \textbf{0.0059} & 15.95 & \textbf{0.0069} & \textbf{18.27} & \textbf{0.0059} & 15.83 \\
\quad Park et al. & 0.0058 & 16.34 & 0.0058 & 16.30 & \textbf{0.0058} & \textbf{16.33} & 0.0058 & 16.73 & 0.0061 & 15.92 & \textbf{0.0058} & \textbf{15.83} & 0.0053 & \textbf{16.04} & 0.0057 & 15.57 & 0.0056 & \textbf{16.14} \\
\addlinespace[3pt]
\cmidrule{1-19}
\addlinespace[3pt]
\multicolumn{19}{@{}l}{\textit{mid-WTP heavy (3,6,3)}} \\\\
\quad Static & 0.0032 & 8.93 & 0.0017 & 5.04 & 0.0012 & 3.24 & 0.0033 & 8.90 & 0.0022 & 5.48 & 0.0012 & 3.38 & 0.0032 & 8.78 & 0.0020 & 5.01 & 0.0011 & 3.14 \\
\quad PRICE (linear) & 0.0055 & 15.68 & 0.0073 & 20.44 & 0.0064 & 18.53 & 0.0059 & 16.99 & 0.0072 & 20.40 & 0.0064 & 18.40 & 0.0058 & 16.20 & \textbf{0.0078} & \textbf{21.26} & \textbf{0.0065} & 18.66 \\
\quad PRICE (exp) & 0.0054 & 14.92 & \textbf{0.0077} & \textbf{21.25} & \textbf{0.0066} & \textbf{19.23} & 0.0062 & \textbf{17.19} & \textbf{0.0073} & 20.29 & 0.0064 & 18.91 & 0.0055 & 15.10 & 0.0077 & 20.45 & 0.0065 & 18.75 \\
\quad PRICE (cubic) & \textbf{0.0059} & 15.89 & 0.0075 & 20.75 & 0.0064 & 18.24 & \textbf{0.0064} & 16.60 & 0.0071 & \textbf{20.42} & \textbf{0.0067} & \textbf{19.20} & 0.0057 & 15.92 & 0.0072 & 20.49 & 0.0065 & \textbf{18.83} \\
\quad Park et al. & \textbf{0.0059} & \textbf{16.21} & 0.0057 & 15.94 & 0.0063 & 16.57 & 0.0063 & 15.91 & 0.0058 & 15.96 & 0.0058 & 15.63 & \textbf{0.0061} & \textbf{16.50} & 0.0062 & 16.41 & 0.0061 & 16.49 \\
\addlinespace[3pt]
\cmidrule{1-19}
\addlinespace[3pt]
\multicolumn{19}{@{}l}{\textit{balanced (4,4,4)}} \\\\
\quad Static & 0.0037 & 9.55 & 0.0019 & 5.27 & 0.0014 & 3.48 & 0.0033 & 9.14 & 0.0020 & 5.08 & 0.0011 & 3.23 & 0.0033 & 9.30 & 0.0020 & 5.02 & 0.0014 & 3.45 \\
\quad PRICE (linear) & 0.0060 & 16.56 & 0.0072 & \textbf{20.58} & 0.0064 & 18.18 & \textbf{0.0066} & \textbf{17.72} & \textbf{0.0073} & \textbf{20.74} & 0.0063 & 18.43 & 0.0059 & 16.54 & \textbf{0.0079} & \textbf{20.42} & 0.0065 & \textbf{18.15} \\
\quad PRICE (exp) & \textbf{0.0061} & 16.43 & \textbf{0.0073} & 20.45 & 0.0063 & 18.63 & 0.0061 & 16.91 & 0.0071 & 20.35 & \textbf{0.0064} & 18.28 & \textbf{0.0065} & \textbf{17.70} & 0.0075 & 19.95 & \textbf{0.0066} & 17.97 \\
\quad PRICE (cubic) & 0.0060 & \textbf{16.82} & 0.0072 & 20.34 & \textbf{0.0064} & \textbf{18.73} & 0.0063 & 16.98 & 0.0071 & 20.41 & 0.0063 & \textbf{18.51} & 0.0063 & 16.56 & 0.0073 & 20.20 & 0.0064 & 17.98 \\
\quad Park et al. & 0.0057 & 16.01 & 0.0057 & 16.01 & 0.0057 & 16.55 & 0.0059 & 15.98 & 0.0059 & 16.36 & 0.0058 & 16.16 & 0.0055 & 15.78 & 0.0059 & 15.64 & 0.0059 & 15.49 \\
\addlinespace[3pt]
\cmidrule{1-19}
\addlinespace[3pt]
\multicolumn{19}{@{}l}{\textit{high-WTP heavy (6,3,3)}} \\\\
\quad Static & 0.0034 & 9.36 & 0.0019 & 5.26 & 0.0012 & 3.37 & 0.0033 & 9.35 & 0.0020 & 5.09 & 0.0012 & 3.41 & 0.0033 & 8.68 & 0.0019 & 5.26 & 0.0011 & 3.14 \\
\quad PRICE (linear) & 0.0058 & 15.83 & 0.0081 & 22.42 & 0.0073 & 20.82 & 0.0059 & 16.50 & \textbf{0.0080} & \textbf{23.04} & 0.0075 & 21.11 & \textbf{0.0060} & 16.35 & 0.0081 & 22.04 & \textbf{0.0077} & \textbf{20.75} \\
\quad PRICE (exp) & \textbf{0.0061} & \textbf{16.75} & 0.0079 & 22.60 & \textbf{0.0075} & \textbf{20.84} & 0.0061 & 16.50 & 0.0078 & 22.11 & 0.0072 & 20.81 & 0.0056 & 15.03 & 0.0080 & \textbf{22.29} & 0.0075 & 20.30 \\
\quad PRICE (cubic) & 0.0060 & 16.62 & \textbf{0.0081} & \textbf{22.74} & 0.0072 & 20.56 & 0.0061 & \textbf{16.97} & 0.0079 & 22.48 & \textbf{0.0076} & \textbf{21.24} & 0.0058 & \textbf{16.58} & \textbf{0.0081} & 21.99 & 0.0076 & 20.37 \\
\quad Park et al. & 0.0060 & 16.26 & 0.0058 & 15.94 & 0.0058 & 16.13 & \textbf{0.0065} & 16.20 & 0.0062 & 15.95 & 0.0058 & 16.30 & 0.0053 & 14.86 & 0.0058 & 16.01 & 0.0065 & 17.27 \\
\bottomrule
\end{tabular}}
\caption{Steady-state revenue rate [\$/s] and total revenue [\$] per method, WTP mixture, and task-duration distribution. \(d = 30\,\mathrm{s}\), \(\mathrm{CAP} \in \{1.0,1.5,2.0\}\), window \(w = 600\,\mathrm{s}\).}
\label{tab:rq3_revenue_d30}
\end{table*}

Table~\ref{tab:rq3_revenue_d30} reports the corresponding provider revenue. Across all \gls{WTP} compositions and duration distributions, all three PRICE pricing functions generate substantially higher revenue rates and totals than Static. The revenue advantage is most pronounced in the high-WTP-heavy configuration with $\text{CAP}=1.5$, where PRICE (linear) achieves a total revenue of up to \$23.04 compared to about \$5 for Static in the corresponding settings. The CAP value controls the revenue-throughput trade-off: lower caps admit more requests at lower prices, whereas higher caps increase selectivity. In the evaluated settings, the intermediate cap \(\mathrm{CAP}=1.5\) often yields the highest total revenue. Under the mid-WTP-heavy composition and $\text{CAP}=1.5$, PRICE generates revenue consistently in the range of \$20.3--\$21.3, roughly four times the Static baseline. Park et al.\ achieves comparable revenue rates to PRICE in some configurations, but this is driven by accepting fewer requests at a fixed high-QoR footprint rather than by adaptive QoR differentiation, as the acceptance ratios in Table~\ref{tab:rq1_big_merged_d60} confirm. Taken together, these results demonstrate that PRICE substantially improves provider revenue compared to Static and non-QoR-aware mechanisms. The revenue gains are driven by higher acceptance rates rather than by per-WTP quality differentiation, which remains moderate in the aggregated view.

\section{Discussion}\label{sec:discussion}

PRICE demonstrates that utilization-dependent pricing is a sufficient control signal for self-organizing resource management in shared edge environments, without requiring cooperative tenants, multi-round negotiation, or prior knowledge of the workload.

\subsection{Robustness Under Execution Variability}

PRICE maintains stability even when actual resource consumption deviates from declared demand. In our evaluation, task durations follow a stochastic distribution around the configured mean, causing transient deviations between reserved and actually consumed resources. The feedback mechanism corrects the operating point based on current reservations and observed runtime behavior rather than relying solely on nominal task parameters, ensuring convergence toward capacity despite such variability. This robustness extends to imperfect resource declarations in general. In practice, actual CPU and memory consumption may also deviate from declared demand due to input variability or runtime scheduling effects, and the feedback mechanism absorbs these deviations without requiring declaration accuracy.

Future work may further strengthen this robustness through lightweight demand forecasting. Recent studies~\cite{behera_time_2023, pradhan_towards_2024} demonstrate that learning-based and time-series models can accurately estimate utilization trends, which would complement the market mechanism by improving bid realism and reducing transient overload during demand spikes.

\subsection{Design Scope and Limitations}

The three pricing functions evaluated produce comparable throughput and acceptance behavior across all tested configurations. 
This indicates that QoR-aware variant selection is the dominant factor, while $p_{\mathrm{cap}}$ mainly controls selectivity under load by decoupling the price level from the functional form. 
Future work includes finer-grained sensitivity analysis of $\beta$, online cap calibration, and coordinating currently local price signals across managers for cross-node load balancing.

PRICE deliberately rejects non-feasible and non-bid-acceptable requests immediately instead of providing acceptance guarantees through server-side queuing. 
This preserves predictable response times and allows latency- or safety-relevant clients to react without queueing delay, but strict guarantees for high-priority requests require additional mechanisms such as reservation-based control or policy-driven resource pools for mission-critical tasks as envisioned by democratic computing~\cite{muhlhauser_towards_2024}. 
Device mobility and network variability affect end-to-end edge behavior but are outside the allocation mechanism studied here. Similarly, PRICE does not yet model tenant-level fairness constraints or strategic repeated bidding, both of which remain future work.

\section{Related Work}
\label{sec:RelatedWork}

\textbf{QoR-aware execution.}
Approximate computing establishes result quality as a controllable variable for resource efficiency~\cite{mittal_survey_2016,liu_approximate_2020}. Loop perforation~\cite{sidiroglou-douskos_managing_2011} and mobile approximate-computing systems~\cite{moreau_approximate_2015} expose accuracy-resource variants related to PRICE's variant model. MobiDiC~\cite{pandey_mobidic_2016} identifies approximable task segments offline and selects among them at runtime to trade accuracy for reduced energy and latency. MARQ~\cite{gropengieser_marq_2025} extends this to distributed AI microservice chains under multi-criteria constraints. MobiQoR~\cite{Li2017} incorporates discrete QoR levels into mobile edge offloading to jointly minimize energy and response time. Zeta~\cite{he_zeta_2012} schedules partial execution within one service, while Ubora~\cite{kelley_measuring_2015} uses answer quality for online admission control. These systems treat variant selection or quality-aware admission as a service-internal optimization problem and do not address how competing tenants sharing a capacity-constrained node should be admitted and differentiated. PRICE couples the variant model to per-variant bids and a utilization-dependent acceptance price, making variant selection a market outcome rather than a centralized decision.

\textbf{Pricing in MEC and edge systems.}
Dynamic pricing for edge resources has been studied extensively, yet none of the existing work couples pricing to per-request execution quality. Park et al.~\cite{park_real-time_2024} derive a utilization-dependent price from marginal cost integrals over a supply curve and target a market equilibrium near capacity. This is structurally the closest prior work to PRICE: both use rising prices as negative feedback under high utilization. The critical gap is that Park et al. treat each request as a fixed-configuration submission, leaving acceptance or rejection as the only control levers. PRICE applies the same feedback loop across a per-request portfolio of QoR variants, turning contention into a variant-selection signal rather than a binary gate. Baek et al.~\cite{baek_three_2020} compare three pricing schemes for IoT edge environments and analyze their revenue-fairness trade-offs at equilibrium, but without any coupling to execution quality. Tütüncüoğlu et al.~\cite{tutuncuoglu_dynamic_2024} and Huang et al.~\cite{huang_pricing_2024} frame price setting as a sequential decision problem in serverless and server-configuration settings, respectively. Wang et al.~\cite{wang_decentralized_2023} and Zheng and Tan~\cite{zheng_decentralized_2025} study decentralized user responses to dynamic prices via mean-field games and reinforcement learning. Han et al.~\cite{han_pricing-based_2024}, and Liao et al.~\cite{liao_adaptable_2025} model server-side price leadership using Stackelberg games that incorporate user energy and offloading costs. Habiba et al.~\cite{habiba_repeated_2024} and Bahreini et al.~\cite{bahreini_mechanisms_2022} provide formal mechanism-design guarantees through repeated auctions and envy-free allocation rules. In all of these works, price regulates the volume of admitted work while the quality of each request remains fixed. To the best of our knowledge, PRICE is the first mechanism in this line to couple utilization-dependent pricing with per-request execution-quality selection under overload.

\textbf{Prices as feedback and incentive signals.}
Kelly et al.~\cite{kelly_rate_1998} show that shadow prices support proportionally fair allocations and network stability under congestion, motivating utilization-dependent prices as negative feedback in resource-constrained systems. Chi et al.~\cite{chi_fairness-aware_2017} apply this perspective to cloud infrastructure, tying prices to resource utilization to improve both revenue and fairness. Wang et al.~\cite{wang_tackling_2024} use pricing as an incentive mechanism in federated learning to correct client availability bias. PRICE transfers the feedback-pricing idea into an edge runtime that combines it with QoR-variant bids, so that a single continuous signal simultaneously governs admission and quality selection under contention.

\section{Conclusion}\label{sec:conclusion}

This paper presented PRICE, a pricing-based incentive mechanism for QoR-aware resource allocation at the edge. By coupling a utilization-dependent price signal with per-request variant selection, PRICE aligns individual tenant bids with provider-level efficiency objectives without requiring tenant coordination or prior knowledge of demand. As utilization increases, rising prices make resource-intensive variants less likely to be bid-acceptable, causing the provider to select lighter submitted variants and driving the system toward sustained high utilization near capacity.

Evaluation under sustained overload demonstrates that PRICE accepts significantly more requests than static and dynamic pricing baselines, stabilizes CPU utilization at a high controlled level, and uses QoR adaptation as a control dimension under contention. These properties are robust across pricing function families, task-duration distributions, and heterogeneous client WTP profiles, and emerge without explicit fairness constraints, workload models, or coordination protocols.

Beyond the immediate results, PRICE establishes a broader principle. Result-quality flexibility is a powerful and underexplored control dimension for resource management in shared edge environments. Treating QoR as an explicit economic variable allows demand peaks to be absorbed through quality adaptation rather than outright rejection, opening a productive direction for future work on self-organizing, incentive-aware edge infrastructure.
\section*{Remarks}
For the purpose of editing, we used OpenAI GPT-5.5, and Grammarly.

\clearpage
\balance

\renewcommand*{\bibfont}{\footnotesize}
\printbibliography

 
\end{document}